\documentclass[10pt,oneside,onecolumn]{article}  

\usepackage{cuted}

\usepackage[body={18.cm,25.cm}]{geometry}                		
\usepackage{graphicx}			
\usepackage{color}

\usepackage{amsmath,amsfonts,amssymb}
\usepackage{mathtools}
\usepackage{epstopdf}
\usepackage{fancyhdr}
\usepackage{empheq}
\usepackage{epstopdf}
\usepackage{pdfpages}

\usepackage{titlesec}
\usepackage{titletoc}
\renewcommand{\thesection}{\Alph{section}}
\renewcommand{\thesubsection}{\Alph{section}.\arabic{subsection}}
\titleformat{\section}{\fontfamily{put}\selectfont\large\bfseries}{\thesection.}{.5em}{}

\usepackage{enumitem}

\definecolor{xll}{rgb}{0.85,0.44,0.84}
\usepackage{hyperref}
\hypersetup{
  colorlinks=true,
  citecolor=brown,
  linkcolor=teal,
  urlcolor=blue,
  urlcolor=cyan,
  pdfborder={0 0 0.5},
}

\DeclareMathAlphabet{\mathcal}{OMS}{cmsy}{m}{n}
\renewcommand{\d}{\mathrm{d}}

\renewcommand{\v}[1]{\mathbf{#1}}

\renewcommand{\rm}[1]{\mathrm{#1}}

\renewcommand{\today}{\number\month/\number\day/\number\year}

\numberwithin{equation}{section}

\usepackage{orcidlink} 

\usepackage{setspace}

\makeatletter
\renewcommand\@makefntext[1]{%
  \noindent\textcolor{gray}{\@thefnmark.\ #1}%
}
\makeatother

\usepackage{utopia}
\usepackage[T1]{fontenc}
\usepackage[utopia]{mathdesign}

\usepackage{authblk}
\usepackage{cite}

\usepackage{titletoc}
\titlecontents{section}
              [0.9cm]
              {\normalsize\bfseries{\textcolor{teal}{\thecontentslabel}}}
              {\normalsize\makebox[.2cm][l]{}}
              {}
              {\normalsize\titlerule*[.5pc]{$\cdot$}\contentspage
              \hspace*{1.cm}}

\titlecontents{subsection}
              [2.25cm]
              {\textcolor{teal}{\small\thecontentslabel}}
              {\small\makebox[0.5cm][l]{}}
              {}
              {\small\titlerule*[.5pc]{$\cdot$}{\textcolor{gray}{\contentspage
              \hspace*{2.5cm}}       }}  

\titleformat{\section}{\large\bfseries}{\thesection}{.5em}{}
\titleformat{\subsection}{\color{black}\it\normalsize}{\thesubsection}{.5em}{}

\usepackage{tikz,xcolor,hyperref}

\definecolor{lime}{HTML}{A6CE39}
\DeclareRobustCommand{\orcidicon}{
	\begin{tikzpicture}
	\draw[lime, fill=lime] (0,0) 
	circle [radius=0.16] 
	node[white] {{\fontfamily{qag}\selectfont \tiny ID}};
	\draw[white, fill=white] (-0.0625,0.095) 
	circle [radius=0.007];
	\end{tikzpicture}
	\hspace{-2mm}
}
\foreach \x in {A, ..., Z}{%
	\expandafter\xdef\csname orcid\x\endcsname{\noexpand\href{https://orcid.org/\csname orcidauthor\x\endcsname}{\noexpand\orcidicon}}
}

\usepackage{multicol}

\allowdisplaybreaks
\usepackage{stfloats}
\usepackage{microtype}
\usepackage{caption}
\DeclareCaptionFont{captionstretch}{\linespread{0.9}\selectfont}
\usepackage{balance}

\counterwithout{equation}{section}

\begin{document}

\title{\Large\bfseries Neutron Star Equation of State with Nucleon Short-Range Correlations:\\ A Concise Review and Open Issues}

\author[1,2]{\normalsize Bao-Jun Cai\orcidlink{0000-0002-8150-1020}\thanks{\textcolor{gray}{bjcai@fudan.edu.cn}}}
\affil[1]{\small \it
Key Laboratory of Nuclear Physics and Ion-beam Application, Institute of Modern Physics, Fudan University, Shanghai 200433, China
}
\affil[2]{\small\it Shanghai Research Center for Theoretical Nuclear Physics, NSFC and Fudan University, Shanghai 200438, China}

\author[3]{\normalsize Bao-An Li\orcidlink{0000-0001-7997-4817}\thanks{\textcolor{gray}{Bao-An.Li$@$etamu.edu}}}
\affil[3]{\small\it Department of Physics and Astronomy, East Texas A\&M University, Commerce, TX 75429-3011, USA}

\author[1,2]{\normalsize Yu-Gang Ma\orcidlink{0000-0002-0233-9900}\thanks{\textcolor{gray}{mayugang$@$fudan.edu.cn}}}

\date{\small\today}
\maketitle

\vspace{-0.25cm}

\begin{abstract}

\begin{spacing}{.95}
Nucleon short-range correlations (SRCs) and the associated high-momentum tail (HMT) in its momentum distribution $n(k)$ represent a universal feature of strongly interacting Fermi systems. In nuclear matter, SRCs arise primarily from the spin-isospin dependence of the tensor and short-range components of the nucleon-nucleon interaction, leading to a substantial depletion of its Fermi sea and a characteristic $k^{-4}$ tail populated predominantly by isosinglet neutron-proton pairs. These microscopic structures modify both the kinetic and interaction contributions to the Equation of State (EOS) of dense matter and thereby influence a broad range of neutron-star (NS) properties. This short review provides a streamlined overview of how SRC-induced changes in $n(k)$ reshape the kinetic EOS, including its symmetry energy part and how these effects propagate into macroscopic NS observables, including mass-radius relations, tidal deformabilities, direct Urca thresholds and core-crust transition. We summarize key existing results, highlight current observational constraints relevant for testing SRC-HMT effects, and outline open questions for future theoretical, experimental, and multimessenger studies of dense nucleonic matter.

\vspace{0.25cm}


\end{spacing}

\end{abstract}

\renewcommand{\contentsname}{\null}
\vspace{-2.cm}
{
\begin{spacing}{1.}
\tableofcontents
\end{spacing}
}

\section{Characteristics of single-nucleon momentum distribution $n_{\v{k}}^J\equiv n(k)$ with an SRC-induced HMT}\label{SEC_INTRO}

\indent

The Equation of State (EOS) of dense nucleonic matter is central to modern nuclear physics and astrophysics\cite{Walecka1974,Collins1975,Chin1977,Freedman1977-1,Freedman1977-2,Freedman1977-3,Baluni1978,Wiringa1988,Akmal1998,Migdal1978,Shuryak1980,Bailin1984,Lattimer2001,Dan02,Steiner2005,LCK08,Alford2008,Watts2016,Ozel2016,Oertel2017,Vidana2018,Bur2021,Dri2021,Sor2024,Kumar2024,Baym2018,Bai2019,Ors2019,Li2019,Dex2021,Lattimer:2021emm,ChenJH24,Shou24,Ding24}. In particular, observables of neutron stars (NSs) reveal the EOS under extreme conditions including high densities and isospin asymmetries as well as strong-field gravity\cite{Shapiro1983,Haen2007,CaiLi25-IPADTOV}, where nuclear interactions become strongly correlated and naturally nonperturbative\cite{Bram14QCD}. Among the microscopic ingredients shaping the EOS, nucleon short-range correlations (SRCs) and the associated high-momentum tail (HMT) in its momentum distribution function $n(k)$ have emerged as essential, experimentally established universal features of strongly interacting Fermi systems\cite{Gio08RMP,Blo08RMP}.

\renewcommand*\figurename{\small FIG.}
\begin{figure}[h!]
\centering
  \includegraphics[width=8.5cm]{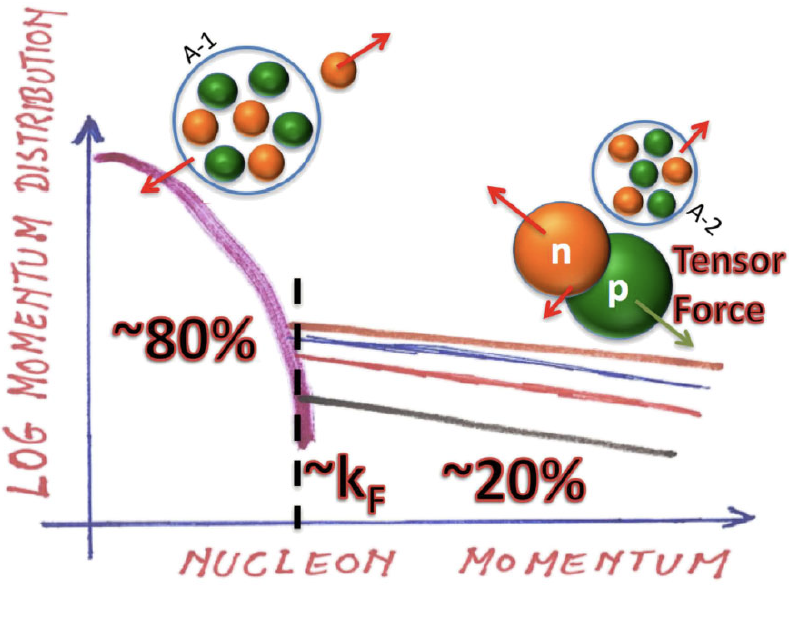}
  \caption{(Color Online). A schematic illustration of the single-nucleon momentum distribution $n_{\mathbf{k}}=n(k)$ in finite nuclei. Below the Fermi momentum $k_{\rm F}$, nucleons predominantly occupy mean-field states, while above the Fermi surface, correlated neutron-proton pairs emerge due to the tensor component of the nuclear force. Figure adapted from Ref.\cite{Hen17RMP}.}
  \label{fig_Hen17}
\end{figure}

SRCs originate from the strong short-range repulsion and tensor components of the nuclear force\cite{bethe}. They deplete the Fermi sea and generate a universal $n(k)\sim k^{-4}$ tail at large $k$\cite{Tan08-a,Tan08-b,Tan08-c}, confirmed in nuclei\cite{Hen14,Sub08}, nuclear matter\cite{Ben93}, and cold atoms\cite{Gio08RMP,Blo08RMP}. The tensor-dominated isosinglet nature of neutron-proton SRC pairs\cite{Hen14} induces a characteristic isospin inversion in asymmetric nuclear matter (ANM): majority nucleons dominate below $k_{\rm F}$, while minority nucleons are overrepresented relatively in the HMT\cite{Hen14}. These correlated pairs possess large relative momentum and small center-of-mass momentum, predominantly back-to-back\cite{Pias23} as sketched in FIG.\,\ref{fig_Hen17} and encode strong spin-isospin dependence tied to in-medium pion and $\rho$-meson dynamics\cite{Ericson1988,Brown:1990kj,Boffi96}. In addition, SRC-HMT physics is known to influence the EMC effect\cite{EMC83,Weinstein:2010rt,CLAS:2019vsb}, further demonstrating its wide relevance to nuclear medium modifications of nucleon structure functions.
Because the EOS depends sensitively on the nucleon momentum distribution, SRC-induced modifications of $n(k)$ directly affect the kinetic contribution to the EOS and the symmetry energy\cite{Li15PRC}:
\begin{equation}\label{def-kinEOS1}
\text{kinetic EOS: }E^{\rm{kin}}\sim\int\frac{\v{k}^2}{2M_{\rm N}}n_{\v k}^J\d\v{k}\sim\int \d kk^4n_{\v k}^J=\int\d k k^4n(k),
\end{equation}
here $J=\rm{n/p}$ is the isospin index for neutrons/protons and $M_{\rm N}\approx939\,\rm{MeV}$ is the static nucleon mass. These changes alter the pressure of neutron-rich matter, the parabolic approximation of the ANM EOS\cite{Bom91,Cai12PRC-S4,Seif14PRC-S4,Gonz17PRC-S4,Pu17PRC-S4}, nucleon effective masses\cite{LCCX18}, single-particle properties\cite{Ding16PRC}, and the composition of dense matter in NSs. Thus, SRCs and the HMT are not simply subleading corrections; they are structural components of a realistic, microscopically grounded EOS.

\begin{figure}[h!]
\centering
\includegraphics[height=8.5cm]{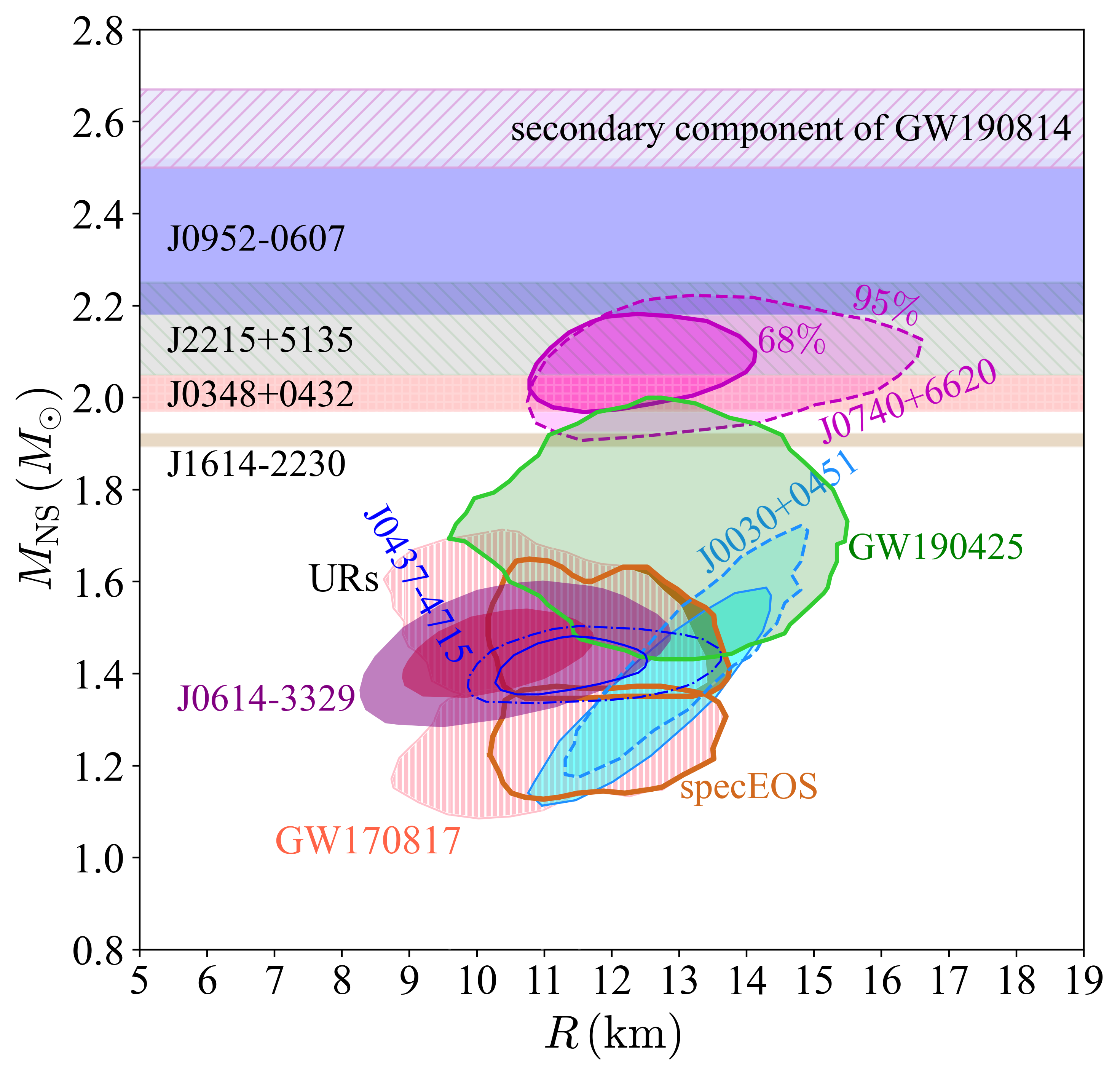}
\caption{(Color Online). A summary on NSs observed, these include the three GW events GW170817\cite{Abbott2017,Abbott2018}, GW190425\cite{Abbott2020-a} and GW190814\cite{Abbott2020}, the NICER mass-radius joint observations for PSR J0740+6620\cite{Riley21}, PSR J0030+0451\cite{Riley19}, PSR J0437-4715\cite{Choud24} and PSR J0614-3329\cite{Mauv25}, the joint X-ray and optical study of redback pulsar PSR J2215+5135\cite{Sull24}, the black widow pulsar PSR 0952-0607\cite{Romani22}, the mass of PSR J1614-2230\cite{Dem10,Arz18} using Shapiro delay and mass of PSR J0348+0432\cite{Ant13} via its spectroscopy.}\label{fig_NSMR-REV}
\end{figure}

The importance of these effects becomes amplified in NSs due to their large isospin asymmetry and high baryon densities. Consequently, the SRC-modified EOS influences the pressure-energy density relation underlying the NS mass-radius (M-R) curves, tidal deformabilities, and the crust-core transition\cite{CaiLi25-IPADTOV}. FIG.\,\ref{fig_NSMR-REV} summarizes current observational M-R constraints for several pulsars, these include the three GW events GW170817\cite{Abbott2017,Abbott2018}, GW190425\cite{Abbott2020-a} and GW190814\cite{Abbott2020}, the NICER mass-radius joint observations for PSR J0740+6620\cite{Riley21} (see also Refs.\cite{Fon21,Miller21,Salmi22,Salmi24,Ditt24}), PSR J0030+0451\cite{Riley19} (see also Refs.\cite{Miller19,Vin24}), PSR J0437-4715\cite{Choud24,Reard24} as well as the recent announced PSR J0614-3329\cite{Mauv25}, the joint X-ray and optical study of redback pulsar PSR J2215+5135\cite{Sull24}, the black widow pulsar PSR 0952-0607\cite{Romani22}, the mass of PSR J1614-2230\cite{Dem10,Arz18} using Shapiro delay and mass of PSR J0348+0432\cite{Ant13} via its spectroscopy; these observational data may guide how the SRC-HMT effects influence NS properties. The redistributed proton fraction associated with the HMT may shift the threshold for direct Urca cooling processes\cite{Yak01,Dong16ApJ,Sed24PRL}, while the altered single-particle spectrum and effective masses affect neutrino opacities, superfluid pairing gaps, thermal evolution, and transport coefficients\cite{Haen2007}. At sub-saturation densities, SRC-driven modifications in $n(k)$ may influence the structure of neutron-rich clusters and nuclear pasta phases; at supra-saturation densities, they affect the stiffness of the EOS and thus the maximum NS mass, which is now tightly constrained about $\gtrsim 2.2\,M_\odot$\cite{CaiLi25-IPADTOV} by astrophysical observations\cite{Riley19,Miller19,Riley21,Miller21,Fon21,Choud24,Reard24,Ditt24,Salmi22,Salmi24,Vin24,Ant13,Arz18,Dem10,Sull24,Romani22,Mauv25}. When combined with multimessenger constraints from NS binary-merger events\cite{Abbott2017,Abbott2018,Abbott2020-a,Abbott2020}, SRC-HMT physics provides a microscopic mechanism linking terrestrial nuclear measurements, many-body theory, and astrophysical probes of dense matter.

The aim of this short review is to highlight how SRCs and the HMT in $n(k)$ reshape the EOS of dense nucleonic matter and how these modifications propagate to NS properties. The review is organized as follows: the remainder of this section briefly discusses the structure of $n(k)$ with an SRC-induced HMT. Section \ref{SEC_EOS} provides a concise overview of SRC-HMT effects on the dense-matter EOS; Section \ref{SEC_NS} presents selected results on SRC-HMT impacts on NS properties; we also highlight several open questions for future studies. Finally, Section \ref{SEC_CC} provides our concluding remarks.
This short review mainly focuses on NS-related issues, and for broader discussions of SRC and HMT physics, the reader is referred to existing reviews, e.g., see Refs.\cite{Hen17RMP,Dick04,Fa17,Arr12,Arr21,Arr22,Dal22,Aum21,Fra81,Fra88,Att15,Fra08,CLMRev25}. Given the limitations of the present authors' knowledge, this brief review is necessarily incomplete and may reflect certain biases in its coverage. Nevertheless, it is hoped that the materials presented here will provide a useful contribution to this rapidly evolving and fascinating field, in which many important and intriguing questions remain to be explored.

\begin{figure}[h!]
\centering
\includegraphics[height=5.5cm]{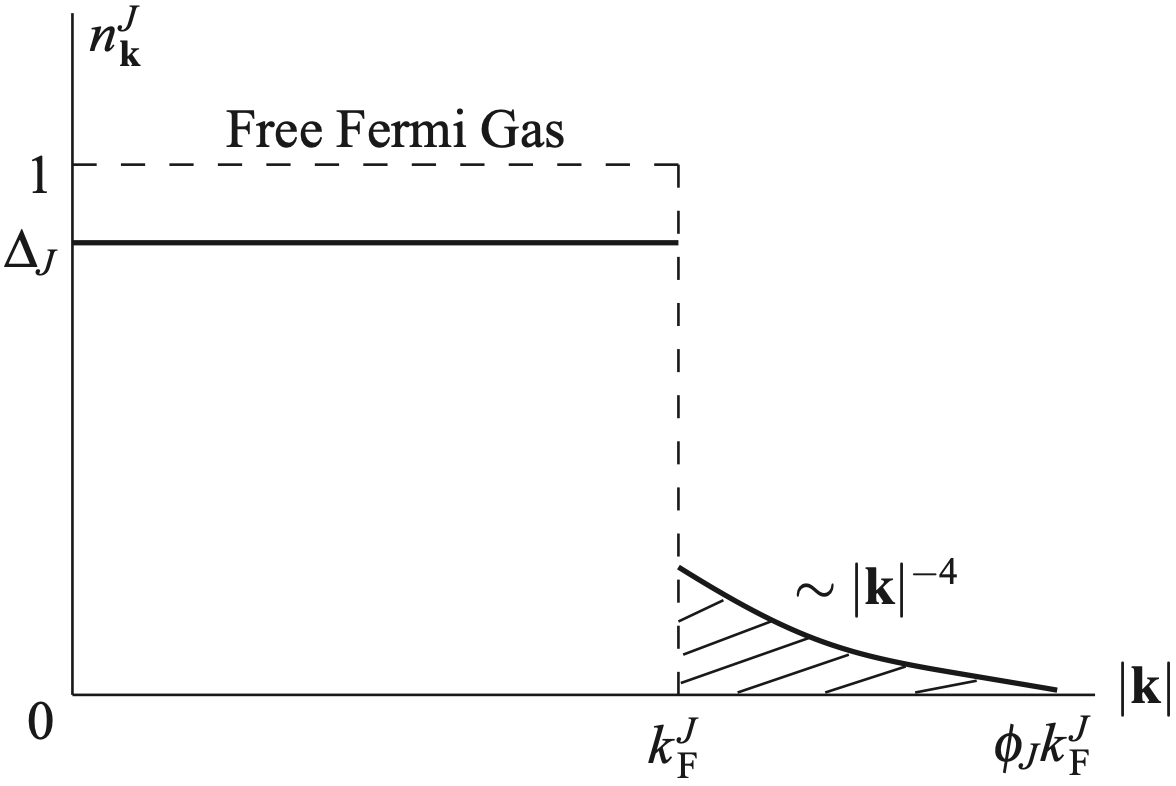}\qquad
\includegraphics[height=5.5cm]{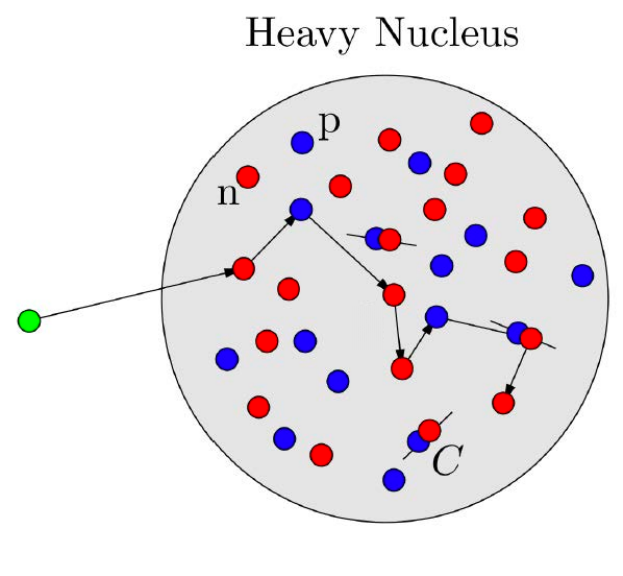}
\caption{(Color Online). Left: nucleon momentum distribution $n_{\mathbf{k}}^{J}$ in isospin ANM, showing the SRC-induced HMT above the Fermi surface and corresponding depletion below it. Figure taken from Ref.\cite{Cai16c}. Right: the back-to-back configuration nature of high-momentum isosinglet neutron-proton pairs, which plays an essential role in shaping the EOS of dense matter and influencing related dynamical processes induced by the incoming projectile (green). Here, ``$C$'' characterizes the contact strength between an np pair.
}\label{fig_nkss}
\end{figure}

The central feature of the single-nucleon momentum distribution $n_{\mathbf{k}}=n(k)$ in the presence of SRCs and tensor correlation is the emergence of a universal HMT of approximate $k^{-4}$ form for $k_{\rm F}\lesssim k\lesssim 2k_{\rm F}$, accompanied by a depletion below the Fermi surface. A widely used parametrization is\cite{Cai15a,Cai16c}
\begin{equation}\label{MDGen}
\boxed{
n_{\mathbf{k}}^{J}(\rho,\delta)=
\begin{cases}
\Delta_{J}, & 0<|\mathbf{k}|<k_{\rm F}^{J},\\
C_{J}\left(\dfrac{k_{\rm F}^{J}}{|\mathbf{k}|}\right)^{4}, & k_{\rm F}^{J}<|\mathbf{k}|<\phi_{J}k_{\rm F}^{J},
\end{cases}}
\end{equation}
as illustrated in the left panel of FIG.\,\ref{fig_nkss}.
Here, $k_{\rm F}^J=k_{\rm F}(1+\tau_3^J\delta)^{1/3}$ is the nucleon Fermi momentum with $k_{\rm F}=(3\pi^2\rho/2)^{1/3}$, as well as $\tau_{3}^{\rm p}=-1$ and $\tau_{3}^{\rm n}=+1$.
The combined effects of SRCs and the tensor component of the nucleon-nucleon interaction give rise to back-to-back neutron-proton pairs (as shown in FIG.\,\ref{fig_Hen17}), which in turn can substantially modify the EOS of dense matter and influence the associated dynamical processes induced by a projectile (green), see the right panel of FIG.\,\ref{fig_nkss}.
The corresponding high-momentum fraction of nucleon $J$ in ANM, according to Eq.\,(\ref{MDGen}), is then given by\cite{Cai15a,CaiLi16a,Cai16b,Cai16c}
\begin{equation}
x_{J}^{\rm{HMT}}=1-\Delta_{J}=3C_{J}\left(1-\phi_{J}^{-1}\right).
\end{equation}
Here $\Delta_{J}$ denotes the zero-momentum depletion, $C_{J}$ the contact coefficient, and $\phi_{J}$ the effective HMT cutoff. Each parameter follows the generic isospin structure $Y_{J}=Y_{0}(1+\tau_{3}^{J}Y_{1}\delta)$\cite{Cai15a}. These coefficients are constrained by microscopic many-body calculations as well as analyses of electron- and proton-nucleus scattering experiments\cite{Migdal57,Lut60,Mah85,Jam89,Bel61,Czy60,Sar80,Ant88,Wei15,Hen15a,Hen14,Tan08-a,Tan08-b,Tan08-c,Hen15b}. Typical values include $C_{0}\approx 0.161$, $C_{1}\approx -0.25$, $\phi_{0}\approx 2.38$, and $\phi_{1}\approx -0.56$\cite{Cai15a}. The discontinuity at the Fermi surface, $Z_{\rm F}^J=\Delta_{J}-C_{J}$, equals the inverse nucleon effective E-mass, as dictated by the Migdal--Luttinger theorem\cite{Migdal57,Lut60,LCCX18,CaiLi16a,Mah85}.
SRC-induced HMTs arise from complex in-medium nucleon-nucleon interactions, including the tensor force, and are probed not only in traditional electron- and proton-nucleus scattering\cite{Boffi96} but also in nucleus-nucleus reactions\cite{Frois1987}. The latter also includes experiments with rare-isotope beams in inverse kinematics. These measurements robustly support the approximate $k^{-4}$ behavior up to $\sim 2k_{\rm F}$\cite{Hen14}. Understanding the microscopic origin of SRCs, their relationship to the EMC effect, and their broad implications for nuclear and astrophysical systems remains a central objective of ongoing and future experimental programs\cite{Hauenstein02012021,Hen:2025rlk}. Relevant efforts include SRC programs\cite{Tu:2020ymk,Hauenstein:2021zql} at the Electron-Ion Collider (EIC) at BNL\cite{Boer:2011fh}, experiments by the R$^{3}$B Collaboration\cite{SRC-r3b,r3b} at GSI-FAIR in Germany\cite{myref}, and forthcoming campaigns at HIAF and the EicC in China\cite{Ye24,EicC-ref}. In addition, active SRC studies at Jefferson Lab, GSI, JINR, and IMP/Lanzhou continue to produce important results\cite{Kahlbow:2023mtc,2023EPJA...59..188A,2023EPJA...59..205F,JHXu25PRR}.

\begin{figure}[h!]
\centering
  \includegraphics[height=6.5cm]{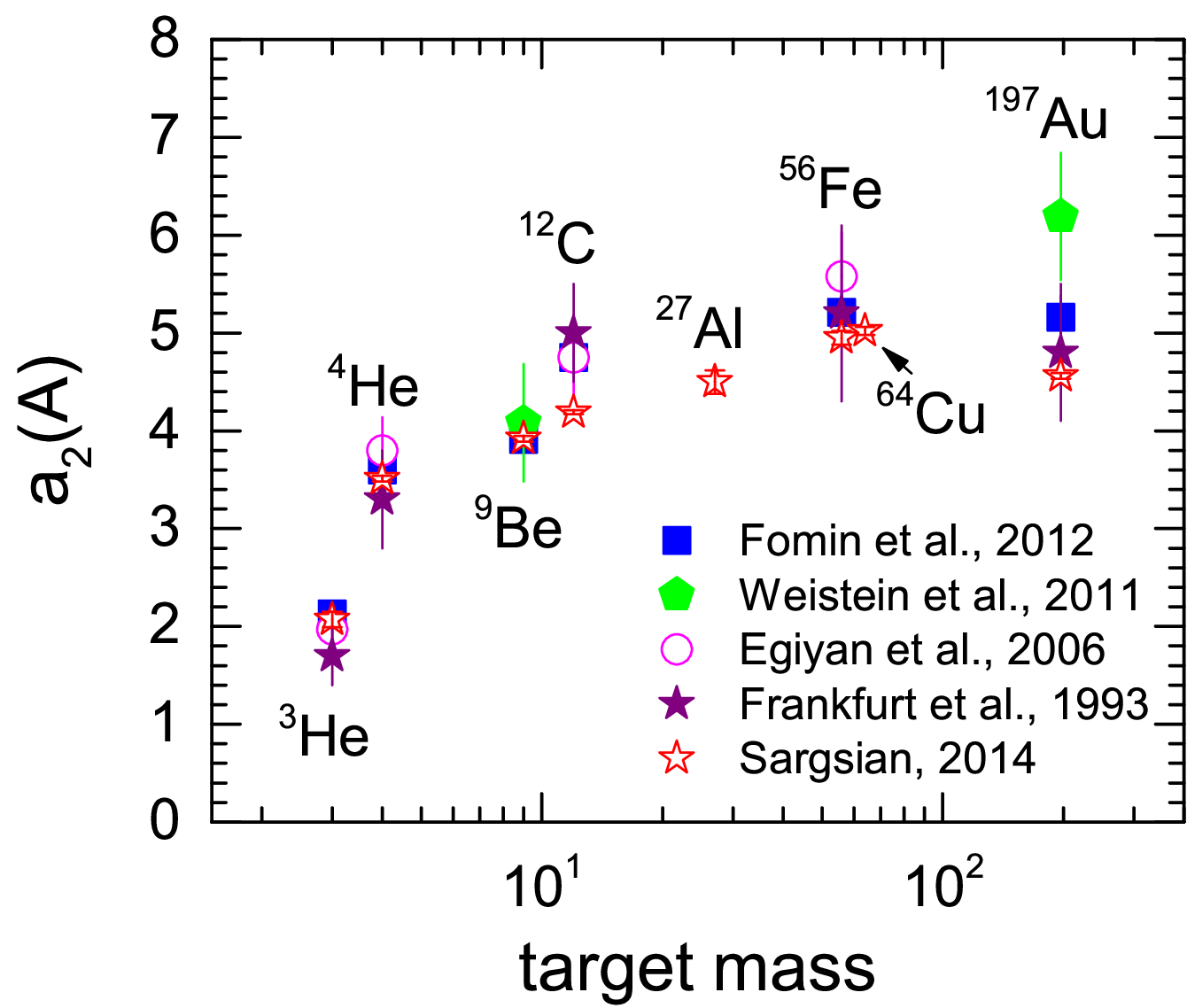}\qquad\qquad
  \includegraphics[height=6.5cm]{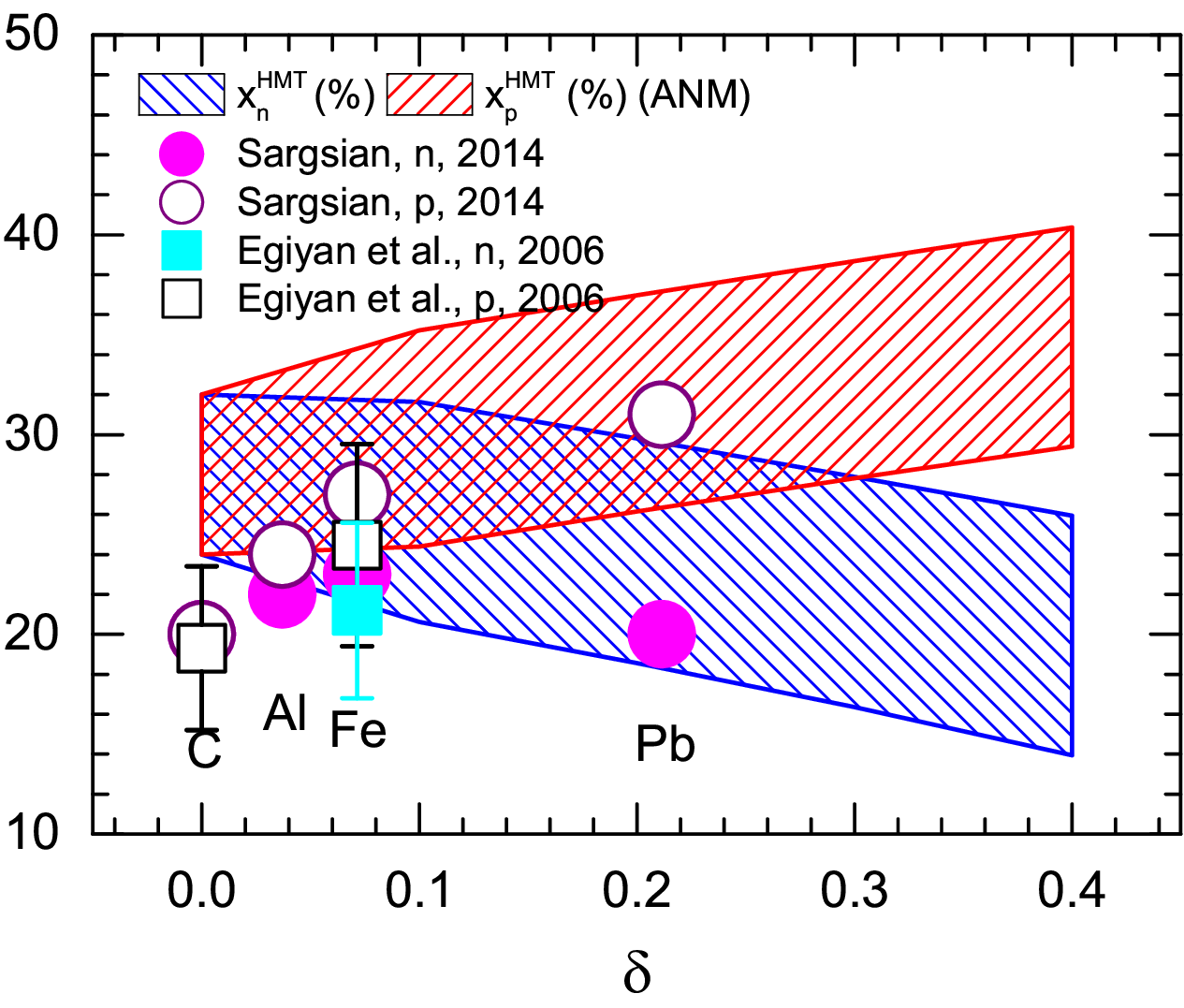}
\caption{(Color Online). Left: experimental $a_2(A)$ for several typical nuclei, i.e., $^3\rm{He}$,
 $^4\rm{He}$, $^9\rm{Be}$, $^{12}\rm{C}$, $^{56}\rm{Fe}$ ($^{63}\rm{Cu}$) and $^{197}\rm{Au}$\cite{Fom12,Wei11,Egi06,Fra93,Sar14}. Right: high-momentum nucleon fractions in heavy nuclei and nuclear matter. Figures adopted from Ref.\cite{LCCX18}.
 }
 \label{fig_a2exp}
\end{figure}

When estimating the SRC-related coefficients such as $C_0$, etc., a key empirical input is the high-momentum nucleon fraction in nuclear matter, which is typically inferred from measurements in finite nuclei. Based on microscopic calculations, the HMT of the nucleon momentum distribution is found to exhibit an approximate scaling from the deuteron to heavier nuclei, $n_{\mathbf k}^{A} = a_2(A)n_{\mathbf k}^{\rm d}$\cite{Fan84,Pie92,Cio96,Fra81,WangR21CPC}. Here $a_2(A)$ represents the relative probability of finding SRC-induced high-momentum nucleon pairs in nucleus $A$ compared with the deuteron\cite{Fra88}. Experimentally, $a_2(A)$ is extracted from the plateau of the per-nucleon inclusive $(\rm{e,e}')$ cross-section ratios in the Bjorken-scaling region $1.5\lesssim x_{\rm{B}} \lesssim 1.9$\cite{Arr12} as $
a_2(A) = (\sigma_A/A)/(\sigma_{\rm d}/2)=2\sigma_A/A\sigma_{\rm d}$.
Representative values for $^3$He, $^4$He, $^{12}$C, $^{56}$Fe, and $^{197}$Au are studied in Refs.\cite{Fom12,Wei11,Egi06,Fra93} as shown in the left panel of FIG.\,\ref{fig_a2exp}, and extrapolation to infinite nuclear matter yields $a_2(\infty)\approx7\pm1$\cite{McG11,Att91,Hen15b,Dai17}.
Moreover, recent developments using the generalized contact formalism (GCF) provide a unified description of inclusive scattering at high $x_{\rm B}$ and momentum transfer $Q^2$, reproducing the observed $a_2$ scaling patterns with realistic interactions\cite{Weiss15PRC,Cruz21NP,Cosyn21PLB,Liang24PLB,Liang24xxx}. These studies highlight that SRC-pair extractions carry $\sim20\%$ uncertainties and require consistent treatments of nuclear structure and relativistic effects. As emphasized in Ref.\cite{Hen17RMP}, apparent inconsistencies in early SRC ratios\cite{Fra93,Egi06} mostly arise from mixing different experimental kinematics. The fundamental physics underlying the well-known linear EMC-SRC correlation\cite{Wei11} remains an open theoretical question\cite{Wang20PRL,LRP2015}. During the last decade, several theoretical frameworks have been benchmarked against the $a_2$ systematics\cite{Rio14,Van12,Van11,Ryck19PRC,Lynn20}. For example, the self-consistent Green's function (SCGF) calculations\cite{Rio14,Dick25B}, using ratios of $\langle n_{\mathbf{k}}^0 / n_{\mathbf{k}}^{\rm d} \rangle_{|\mathbf{k}|=400\text{-}550\,{\rm{MeV}}}$, yield $a_2(\rho)\approx b_1\rho^{b_2}$ with $b_1\approx7\text{-}10$ and $b_2\approx 0.4\text{-}0.5$, but their extrapolated saturation value underestimates $a_2(\infty)$ due in part to a lower high-$k$ cutoff than used experimentally ($k\approx300\text{-}600\,\rm{MeV}$)\cite{Hen14}. Independent-particle-model (IPM) estimates\cite{Van11,Van12} based on the number of isosinglet pn spin-triplet pairs predict an approximately linear rise of $a_2$ up to $A\approx40$, followed by saturation, which is consistent with the data. Similarly, the low-order correlation operator approximation (LCA)\cite{Ryck19PRC} provides alternative estimates using the ratio or integral of $P^A(p)$ at large momenta, yielding total SRC scaling factors $\sim4$-5 with non-negligible (pp, nn) contributions in addition to pn pairs.
On the other hand, Quantum Monte Carlo (QMC) calculations with chiral EFT interactions\cite{Lynn20} successfully reproduce the $a_2$ systematics up to $^{40}$Ca and confirm the empirical EMC-SRC correlation for light nuclei. Their predicted saturation value, $a_2(A\to\infty)\approx4.9\text{-}5.4$, is consistent with measurements on heavy nuclei.
Experimental studies further indicate that reproducing both the deuteron D-state probability ($4$-$5\%$)\cite{Fra88} and $a_2(\infty)$ requires an HMT fraction in symmetric nuclear matter (SNM) of $x_{\rm{SNM}}^{\rm{HMT}}\approx28\%\pm4\%$\cite{Hen14,Egi06,Pia06,Shn07,Wei11,Kor14,Hen15b}. In $^{12}$C, about $20\%$ of nucleons participate in SRCs, $90\%\pm10\%$ of which are pn pairs\cite{Sub08,Egi06}, corresponding to a pn-to-pp enhancement of $\sim20$, a pattern that persists for heavy nuclei\cite{Hen14}. By isospin symmetry, this implies a much smaller HMT fraction in pure neutron matter (PNM), $x_{\rm{PNM}}^{\rm{HMT}}\approx1.5\%\pm0.5\%$. 
Studies on extracting the $a_2$ factor and related issues are given in Refs.\cite{Niu22PRC,Torr18PLB,Pas20,Duer18Nature,Li22Nature, Weiss21PRC,Shang25,Meng23PRC}.

These empirical and theoretical findings indicate a pronounced isospin dependence of the HMT fractions $x_J^{\rm{HMT}}$ in ANM\cite{Cai15a,Hen14,Sar14}, with proton high-momentum fractions exceeding those of neutrons, as illustrated in the right panel of FIG.\,\ref{fig_a2exp}. Recent quasi-elastic $(\rm{p,2p})$ and $(\rm{e,e'p})$ measurements in neutron-rich nuclei further quantify proton-neutron momentum asymmetries\cite{Pas20,Duer18Nature}. Such constraints are essential for determining the coefficients $C_0$ and related parameters in microscopic EOS analyses, and consequently, precise knowledge of $a_2(\infty)$, $x_J^{\rm{HMT}}$, and other SRC parameters is vital for constructing a reliable dense-matter EOS and interpreting multi-messenger observations of NSs.

Remarkably, despite the nearly 25 orders-of-magnitude difference in densities between nuclear matter and ultra-cold atomic Fermi gases, both systems exhibit very similar high-momentum tails with the universal $1/|\v{k}|^4$ scaling. In both cases, short-range two-body correlations govern the large-momentum behavior: the tensor-force-induced np pairs in nuclei and the strong s-wave contact interactions in cold atoms. As a result, the probability of finding correlated pairs, quantified by the contact parameter, shows comparable magnitudes, highlighting an underlying universality in quantum many-body systems across vastly different scales\cite{Hen15a,Gio08RMP,Blo08RMP}.

\section{SRC-HMT Effects on Dense Matter EOS: Concise Overview of Main Findings}\label{SEC_EOS}

\indent

In this section, we briefly discuss the impact of an SRC-induced HMT on the EOS of dense matter, focusing on selected results from the literature; for a more detailed investigation, see Ref.\cite{CLMRev25}.  
As indicated by Eq.\,(\ref{def-kinEOS1}), the modification of $n_{\v k}^J$ due to the SRC-HMT directly affects the kinetic EOS. Specifically, one has
\begin{equation}\label{kinE}
\boxed{
E^{\rm{kin}}(\rho,\delta)=\frac{1}{\rho}\frac{2}{(2\pi)^3}
\sum_{J=\rm{n,p}}\int_0^{\phi_Jk_{\rm{F}}^J}\frac{\v{k}^2}{2M_{\rm N}}n_{\v{k}}^J(\rho,\delta)\d\v{k}, \text{ with respect to }~\frac{2}{(2\pi)^3}\int_0^{\infty}n^J_{\v{k}}(\rho,\delta)\d\v{k}
=\rho_J=\frac{(k_{\rm{F}}^{J})^3}{3\pi^2}.}
\end{equation}
The left-hand side can be expanded as 
$E^{\rm{kin}}(\rho,\delta)\approx E_0^{\rm{kin}}(\rho)+E_{\rm{sym}}^{\rm{kin}}(\rho)\delta^2+E_{\rm{sym,4}}^{\rm{kin}}(\rho)\delta^4+\cdots$, defining respectively the kinetic EOS of SNM, the kinetic symmetry energy, and the fourth-order kinetic symmetry energy from left to right\cite{Cai15a}.  
Using the $n_{\v k}^J$ of Eq.\,(\ref{MDGen}), the first two terms are obtained as\cite{Cai15a}
\begin{align}
E^{\rm{kin}}_0(\rho)=&\frac{3}{5}E_{\rm{F}}(\rho)\left[
1+{C}_0\left(5\phi_0+\frac{3}{\phi_0}-8\right)\right]
\equiv \frac{3}{5}E_{\rm{F}}(\rho)\left[
1+C_0\Phi_0\right],\label{E0kin}\\
E_{\rm{sym}}^{\rm{kin}}(\rho)=&\frac{1}{3}E_{\rm{F}}(\rho)\Bigg[1+{C}_0\left(1+3{C}_1\right)\left(5\phi_0+\frac{3}{\phi_0}-8\right)
+3{C}_0\phi_1\left(1+\frac{3}{5}{C}_1\right)\left(5\phi_0-\frac{3}{\phi_0}\right)+\frac{27{C}_0\phi_1^2}{5\phi_0}\Bigg],\label{Esymkin}
\end{align}
where $E_{\rm F}(\rho)=k_{\rm F}^2/2M_{\rm N}$ is the nucleon Fermi energy in SNM. The second equality in Eq.\,(\ref{E0kin}) defines the factor $\Phi_0$, which is greater than 1 due to $\phi_0>1$.  
The SRC-modified kinetic EOS has two main consequences\cite{Cai15a}:
\begin{enumerate}[label=(\alph*),leftmargin=*]
    \item The factor inside the brackets of Eq.\,(\ref{Esymkin}) is negative, indicating that the kinetic symmetry energy is significantly reduced by SRC-HMT, despite the enhancement of the kinetic EOS of SNM.
    \item The full EOS of ANM $E(\rho,\delta)=E^{\rm{kin}}(\rho,\delta)+E^{\rm{pot}}(\rho,\delta)$ generally includes a potential contribution $E^{\rm{pot}}(\rho,\delta)$. To maintain consistency with empirical constraints around saturation density\cite{LCXZ2021}, the potential part (either of SNM or of the symmetry energy) must readjust correspondingly. This leads to changes in the density dependence of the EOS, which can have sizable effects on NS properties.
Similarly, Eq.\,(\ref{kinE}) can be generalized in two ways:
\begin{enumerate}[label=(\arabic*),leftmargin=*]
    \item Including relativistic corrections to the nucleon kinetic energy gives\cite{CaiLi22PRCFFG}
        \begin{equation}\label{kinE-R}
E^{\rm{kin}}(\rho,\delta)=\frac{1}{\rho}\frac{2}{(2\pi)^3}
\sum_{J=\rm{n,p}}\int_0^{\phi_Jk_{\rm{F}}^J}\frac{\v{k}^2}{2M_{\rm N}}\left(1-\frac{\v{k}^2}{4M_{\rm N}^2}+\frac{\v{k}^4}{8M_{\rm N}^4}-\frac{5\v{k}^6}{64M_{\rm N}^6}+\cdots\right)n_{\v{k}}^J(\rho,\delta)\d\v{k}.
\end{equation}
\item Including effective mass corrections yields\cite{Cai16c}
\begin{equation}
    \label{kinE-Mcorr}
E^{\rm{kin}}(\rho,\delta)=\frac{1}{\rho}\frac{2}{(2\pi)^3}
\sum_{J=\rm{n,p}}\int_0^{\phi_Jk_{\rm{F}}^J}\left(
\sqrt{{\v{k}}^2+M_{\rm N}^{\ast,2}}-M_{\rm N}^{\ast}
\right)n_{\v{k}}^J(\rho,\delta)\d\v{k},
\end{equation}
where $M_{\rm N}^{\ast}$ is some type of the nucleon effective mass\cite{LCCX18}.
\end{enumerate}
In both cases, the potential part should adjust consistently with the kinetic contribution.
\end{enumerate}

More generally, when evaluating any quantity affected by the SRC-HMT, the following prescription applies. For an arbitrary function $f$ at zero temperature, the free Fermi gas (FFG) step function is replaced by the momentum distribution $n_{\v{k}}^J$, which includes both the depletion below $k_{\rm F}^J$ and the HMT:
\begin{equation}
\boxed{
\int_0^{k_{\rm{F}}^J} \overbrace{(\rm{FFG\ step\ function})}^{\Theta(k_{\rm F}^J-|\v{k}|)} f  \d\v{k} 
\longrightarrow 
\int_0^{\phi_J k_{\rm{F}}^J} n_{\v{k}}^J(\rm{HMT})f \d\v{k}.}
\end{equation}
In this sense, $n_{\v{k}}^J$ not only modifies the kinetic contribution through the momentum integration, but also effectively incorporates interaction effects, since coefficients such as $C_0$ inherently reflect nucleon-nucleon interactions\cite{Cai16c}. In other words, the modification should be understood in the quasi-nucleon picture.

\begin{figure}[h!]
\centering
  \includegraphics[height=6.cm]{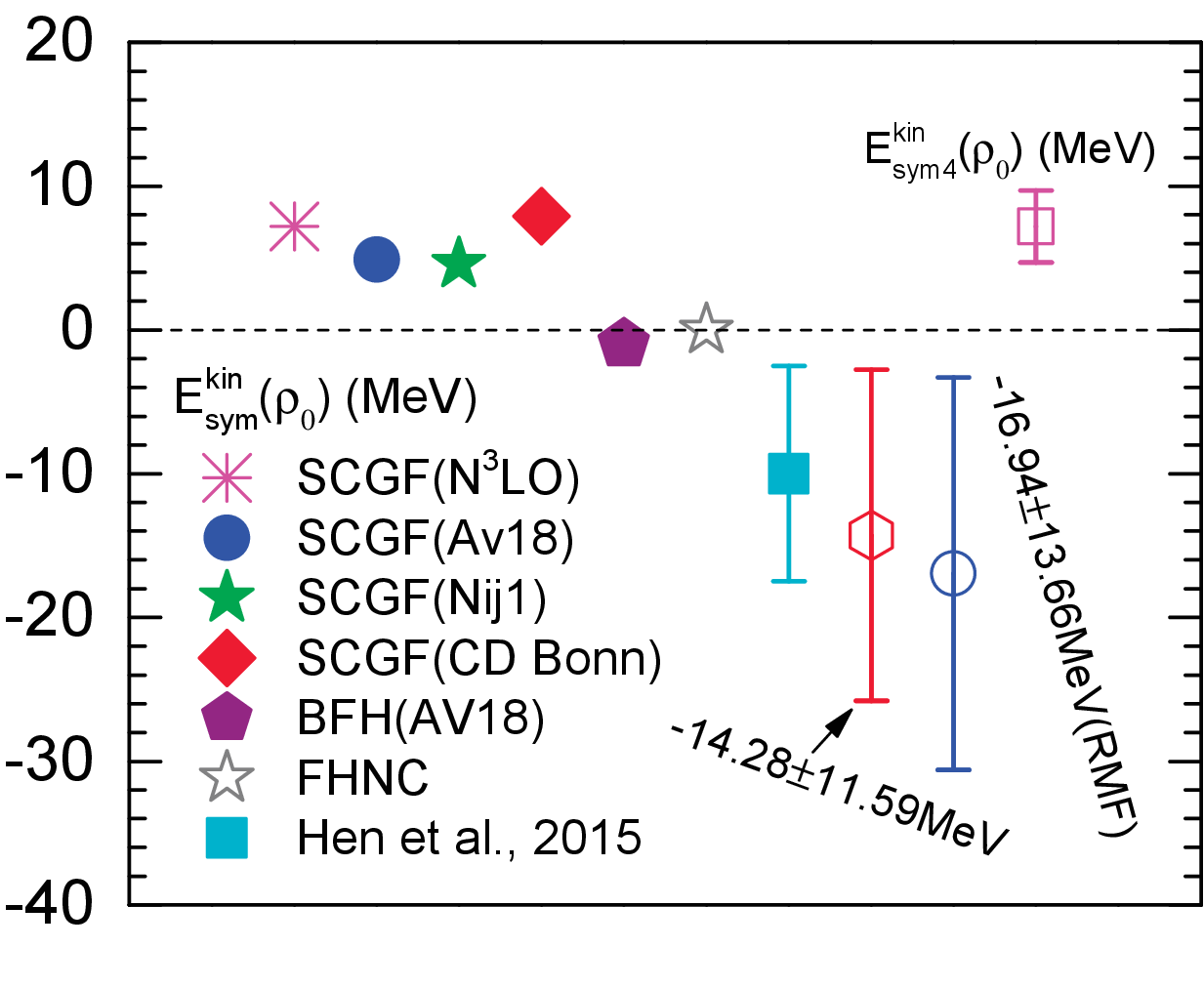}\qquad
  \includegraphics[height=6.cm]{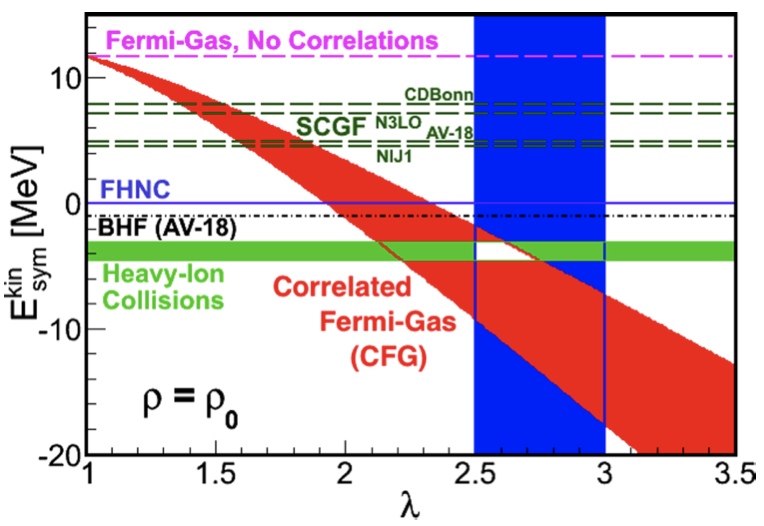}
  \caption{(Color Online). Left: the kinetic symmetry energy and the isospin-quartic term including the SRC-induced HMT in comparison with predictions by other theories. Figure taken from Ref.\cite{LCCX18}. 
  Right: the kinetic symmetry energy using a phenomenological parameterization of Ref.\cite{Hen15b}; here the parameter $\lambda$ is the high-momentum cutoff.
  }
  \label{fig_esym0}
\end{figure}

The SRC-induced reduction of the kinetic symmetry energy relative to the FFG prediction is a robust feature observed across many-body theories that include SRC effects. As illustrated in the left panel of FIG.\,\ref{fig_esym0}, comparisons with microscopic many-body and phenomenological models, including the SCGF prediction (N$^3$LO, AV18, Nij1, CD Bonn)\cite{Car12}, BHF with AV18 plus Urbana IX three-body-force\cite{Vid11}, the Fermi hypernetted chain (FHNC) model\cite{Lov11}, and phenomenological parametrizations\cite{Hen15b}, consistently show that SRC reduces the kinetic symmetry energy, although quantitative differences exist. 
In particular, one finds $E_0^{\rm{kin}}(\rho_0)\approx39.77\pm8.13\,\rm{MeV}$ for the kinetic EOS of SNM, 
$E_{\rm{sym}}^{\rm{kin}}(\rho_0)\approx-14.28\pm11.59\,\rm{MeV}$ for the kinetic symmetry energy, and a large quartic term
$E_{\rm{sym,4}}^{\rm{kin}}(\rho_0)\approx7.18\pm2.52\,\rm{MeV}$\cite{Cai15a}. The latter two quantities are displayed in the left panel of FIG.\,\ref{fig_esym0}.
Including relativistic corrections further gives 
$
E_{\rm{sym}}^{\rm{kin}}(\rho_0)\approx-12.01\pm8.23\,\rm{MeV}
$\cite{CaiLi22PRCFFG}.
In the right panel of FIG.\,\ref{fig_esym0}, the kinetic symmetry energy obtained in Ref.\cite{Hen15b} is shown, where $\lambda$ denotes the high-momentum cutoff. In that work, the single-nucleon momentum distribution is parameterized as
\begin{equation}\label{def_nk1}
\overline{n}(\v{k})=\left\{\begin{array}{ll}A,&
0<|\textbf{k}|<k_{\textrm{F}},\\
{\displaystyle
C_{\infty}}/{\displaystyle|\textbf{k}|^4},&k_{\textrm{F}}<|\textbf{k}|<\lambda
k_{\textrm{F}}^0,
\end{array}\right.
\end{equation}
where $C_{\infty}$ is the HMT strength, $k_{\rm F}^0$ is the Fermi momentum at saturation density $\rho_0$, and $A$ is fixed by the normalization condition 
$[{g}/{(2\pi)^3}]\int_{0}^{\infty} \overline{n}(\v{k})\d\v{k}=1$ with $g=2$ (slightly different from that in Eq.\,(\ref{kinE})). 
Using this distribution, one obtains
\begin{align}
\langle k^j\rangle=\frac{g}{(2\pi)^3}\int_0^{\infty}k^j\overline{n}(\v{k})\d\v{k}
=
\frac{3k_{\rm{F}}^j}{j+3}-\frac{gc_0k_{\rm{F}}^j}{2\pi^2}\left[\frac{4j}{j^2+2j-3}-\frac{1}{\lambda}\frac{3}{j+3}\left(\frac{\rho}{\rho_0}\right)^{1/3}
-\frac{\lambda^{j-1}}{j-1}\left(\frac{\rho_0}{\rho}\right)^{j/3-1/3}\right],
\end{align}
where $C_{\infty}=c_0k_{\rm{F}}$, with $c_0\sim a_2(\infty)$ taken as a constant\cite{Hen15b}.  
For the special case $j=2$, and using the parabolic approximation, the kinetic symmetry energy becomes\cite{Hen15b}
\begin{equation}
E_{\rm{sym}}^{\rm{kin}}(\rho)
=(2^{2/3}-1)\cdot\frac{3k_{\rm{F}}^2}{10M_{\rm N}}
-\frac{c_0}{\pi^2}\frac{k_{\rm{F}}^2}{2M_{\rm N}}\left[\frac{3}{5}\frac{1}{\lambda}\left(\frac{\rho}{\rho_0}\right)^{1/3}
+\lambda\left(\frac{\rho_0}{\rho}\right)^{1/3}-\frac{8}{5}\right],
\end{equation}
where the HMT in PNM is neglected (an approximation).  
This analysis yields $E_{\rm{sym}}^{\rm{kin}}(\rho_0)\approx-10\pm7.5\,\rm{MeV}$\cite{Hen15b}, as shown by the red band in the right panel of FIG.\,\ref{fig_esym0}.
The reduction of the kinetic symmetry energy by SRCs not only impacts our understanding of the origin of symmetry energy but also influences isovector observables\cite{Hen15b,Hen15a,Li15PRC,Yong1,Yong2,Yong3,Yong4,Zhang19,ShenL22,Guo23NPA,Guo24PRC,Guo23PRC,Guo25PRC,Wada25PRC,Reich25,JHXu24PLB,JHXu25PRR}, such as the free neutron/proton and $\pi^-/\pi^+$ ratios in heavy-ion collisions. Furthermore, the potential effects of a large quartic term\cite{Cai15a} on heavy-ion collisions have not been explored due to the relatively small isospin asymmetry $\delta$ in typical reactions. However, in neutron-rich systems, such as reactions involving rare isotopes or peripheral collisions of heavy nuclei with thick neutron skins and NS environments, the isospin quartic term may play an important role. Typically, the EOS of neutron-rich nucleonic matter is expressed as a sum of a FFG kinetic term and a potential energy truncated at the isospin-quadratic term. Results in Ref.\cite{Cai15a} highlight the importance of including the isospin-quartic term in both kinetic and potential contributions. An accurate extraction of the potential isospin-quartic term $E^{\rm{pot}}_{\rm{sym,4}}(\rho)\delta^4$ requires using the kinetic EOS of quasi-particles, which features a reduced symmetry energy and an enhanced quartic term due to the isospin-dependent HMT.  

Two complementary approaches have been developed in the literature to incorporate the SRC-HMT-affected potential contribution into the full EOS of ANM. The first adopts an effective nucleon potential model, while the second embeds SRC-induced HMT effects directly into a relativistic field-theoretical framework. 
As an example of the former, Ref.\cite{CaiLi22Gog} employs a Gogny-type momentum-dependent potential\cite{Chen05,XuJ10,XuJ15}, leading to
\begin{align}\label{EDF}
E(\rho,\delta)=&E^{\rm{kin}}(\rho,\delta)
+\frac{A_\ell(\rho_{\rm{p}}^2+\rho_{\rm{n}}^2)}{2\rho\rho_0}
+\frac{A_{\rm{u}}\rho_{\rm{p}}\rho_{\rm{n}}}{\rho\rho_0}
+\frac{B}{\sigma+1}\left(\frac{\rho}{\rho_0}\right)^{\sigma}(1-x\delta^2)
+\sum_{J,J'}\frac{C_{J,J'}}{\rho\rho_0}\int\d\v{k}\d\v{k}'f_J(\v{r},\v{k})f_{J'}(\v{r},\v{k}')\Omega(\v{k},\v{k}'),
\end{align}
where $E^{\rm{kin}}(\rho,\delta)$ is defined in Eq.\,(\ref{kinE}), $\Omega$ is the regulator function and $f_J(\v{r},\v{k})=(2/h^3)n_{\v{k}}^J(\rho,\delta)$ is the nucleon phase-space distribution function. 
In the FFG limit, $f_J(\v{r},\v{k})=(1/4\pi^3)\Theta(k_{\rm F}^J-|\v{k}|)$. 
The full EOS of SNM and the corresponding kinetic symmetry energy within this framework are summarized in Ref.\cite{CaiLi22Gog}.
A parallel treatment can be carried out within relativistic models. 
For instance, in the nonlinear Walecka model\cite{Cai16c,Cai12PRC-S4,Serot1986,Muller1996NPA,CY25},
\begin{align}
\mathcal{L}_{\rm{Walecka}}=&\overline{\psi}\left[\gamma_{\mu}\left(i\partial^{\mu}-g_{\omega}\omega^{\mu}-g_{\rho}\vec{\rho}^{\mu}\cdot\vec{\tau}\right)-\left(M_{\rm N}-g_{\sigma}\sigma\right)\right]\psi
+\frac{1}{2}\left(\partial_{\mu}\sigma\partial^{\mu}\sigma-m_{\sigma}^2\sigma^2\right)
-\frac{1}{4}\omega_{\mu\nu}\omega^{\mu\nu}
+\frac{1}{2}m_{\omega}^2\omega_{\mu}\omega^{\mu}
\notag\\
&-\frac{1}{3}b_{\sigma}M_{\rm N}\left(g_{\sigma}\sigma\right)^3
-\frac{1}{4}c_{\sigma}\left(g_{\sigma}\sigma\right)^4
+\frac{1}{4}c_{\omega}\left(g_{\omega}^2\omega_{\mu}\omega^{\mu}\right)^2
+\frac{1}{2}m_{\rho}^2\vec{{\rho}}_{\mu}\cdot\vec{{\rho}}^{\mu}
-\frac{1}{4}\vec{{\rho}}_{\mu\nu}\cdot\vec{{\rho}}^{\mu\nu}
+\frac{1}{2}\Lambda_{\rm V}g_{\rho}^2\vec{\rho}_{\mu}\cdot\vec{\rho}^{\mu}\Lambda_{\rm V}g_{\omega}^2\omega_{\mu}\omega^{\mu},
\label{rmf_lag}
\end{align}
one obtains the full EOS of SNM, $E_0(\rho)=\varepsilon_0/\rho-M_{\rm N}$\cite{Cai16c},
\begin{align}
\varepsilon_0(\rho)=&\frac{\Delta_0M_0^{\ast,4}}{\pi^2}\left[\frac{1}{4}\theta\left(1+\theta^2\right)^{3/2}-\frac{1}{8}\theta\sqrt{1+\theta^2}-\frac{1}{8}\rm{arcsinh}\,\theta\right]
+\frac{C_0k_{\rm{F}}^4}{\pi^2}\left[V(\theta)-\sqrt{1+\frac{1}{\phi_0^2\theta^2}}+\sqrt{1+\frac{1}{\theta^2}}\right]
\notag\\
&
+\frac{1}{2}m_{\sigma}^2\overline{\sigma}^2
+\frac{1}{3}b_{\sigma}M_{\rm N}g_{\sigma}^3\overline{\sigma}^3
+\frac{1}{4}c_{\sigma}g_{\sigma}^4\overline{\sigma}^4
+\frac{1}{2}m_{\omega}^2\overline{\omega}_0^2
+\frac{3}{4}c_{\omega}(g_{\omega}\overline{\omega}_0)^3,
\label{cc2_eps0}
\end{align}
together with the symmetry energy:
\begin{align}
E_{\rm{sym}}^{\rm{kin}}(\rho)=&
\frac{g_{\rho}^2\rho}{2Q_{\rho}}
+\frac{k_{\rm F}^2}{6E_{\rm F}^{\ast}}\left[1-3C_0\left(1-\frac{1}{\phi_0}\right)\right]
-3E_{\rm F}^{\ast}C_0\left[C_1\left(1-\frac{1}{\phi_0}\right)+\frac{\phi_1}{\phi_0}\right]
\notag\\
&
-\frac{9M_0^{\ast,4}}{8k_{\rm F}^3}\frac{C_0\phi_1(C_1-\phi_1)}{\phi_0}
\left[2\theta\left(1+\theta^2\right)^{3/2}-\theta\sqrt{1+\theta^2}-\rm{arcsinh}\,\theta\right]
\notag\\
&
+\frac{2k_{\rm F}C_0(6C_1+1)}{3}\left[V(\theta)-\sqrt{1+\frac{1}{\phi_0^2\theta^2}}+\sqrt{1+\frac{1}{\theta^2}}\right]
+\frac{C_0(4+3C_1)}{3}\left[\frac{F_{\rm F}^{\ast}(1+3\phi_1)}{\phi_0}-E_{\rm F}^{\ast}\right]
\notag\\
&
+\frac{3k_{\rm F}C_0}{2}\Bigg[
\frac{(1+3\phi_1)^2}{9}\left(\frac{\phi_0k_{\rm F}}{F_{\rm F}^{\ast}}
-\frac{2F_{\rm F}^{\ast}}{\phi_0k_{\rm F}} \right)
+\frac{2F_{\rm F}^{\ast}(3\phi_1-1)}{9\phi_0k_{\rm F}}
-\frac{1}{9}\frac{k_{\rm F}}{E_{\rm F}^{\ast}}
+\frac{4E_{\rm F}^{\ast}}{9k_{\rm F}}
\Bigg],
\label{EsymkinHMT}
\end{align}
where $Q_{\rho}=m_{\rho}^2+\Lambda_{\rm V}g_{\omega}^2g_{\rho}^2\overline{\omega}_0^2$, $\theta=k_{\rm F}/M_0^{\ast}$ is generally smaller than 1, and $V(\theta)=\rm{arcsinh}(\phi_0\theta)-\rm{arcsinh}\,\theta$. 
We also define $F_{\rm F}^{\ast}=\sqrt{M_0^{\ast,2}+p_{\rm F}^2}$ with $p_{\rm F}=\phi_0k_{\rm F}$\cite{Cai16c}. 
Eq.\,(\ref{EsymkinHMT}) reduces to the standard RMF expression $E_{\rm{sym}}(\rho)=k_{\rm F}^2/6E_{\rm F}^{\ast}+{g_{\rho}^2\rho}/{2Q_{\rho}}$ when $\phi_0=1$ and $\phi_1=0$ are taken.

\begin{figure}[h!]
\centering
  \includegraphics[width=7.5cm]{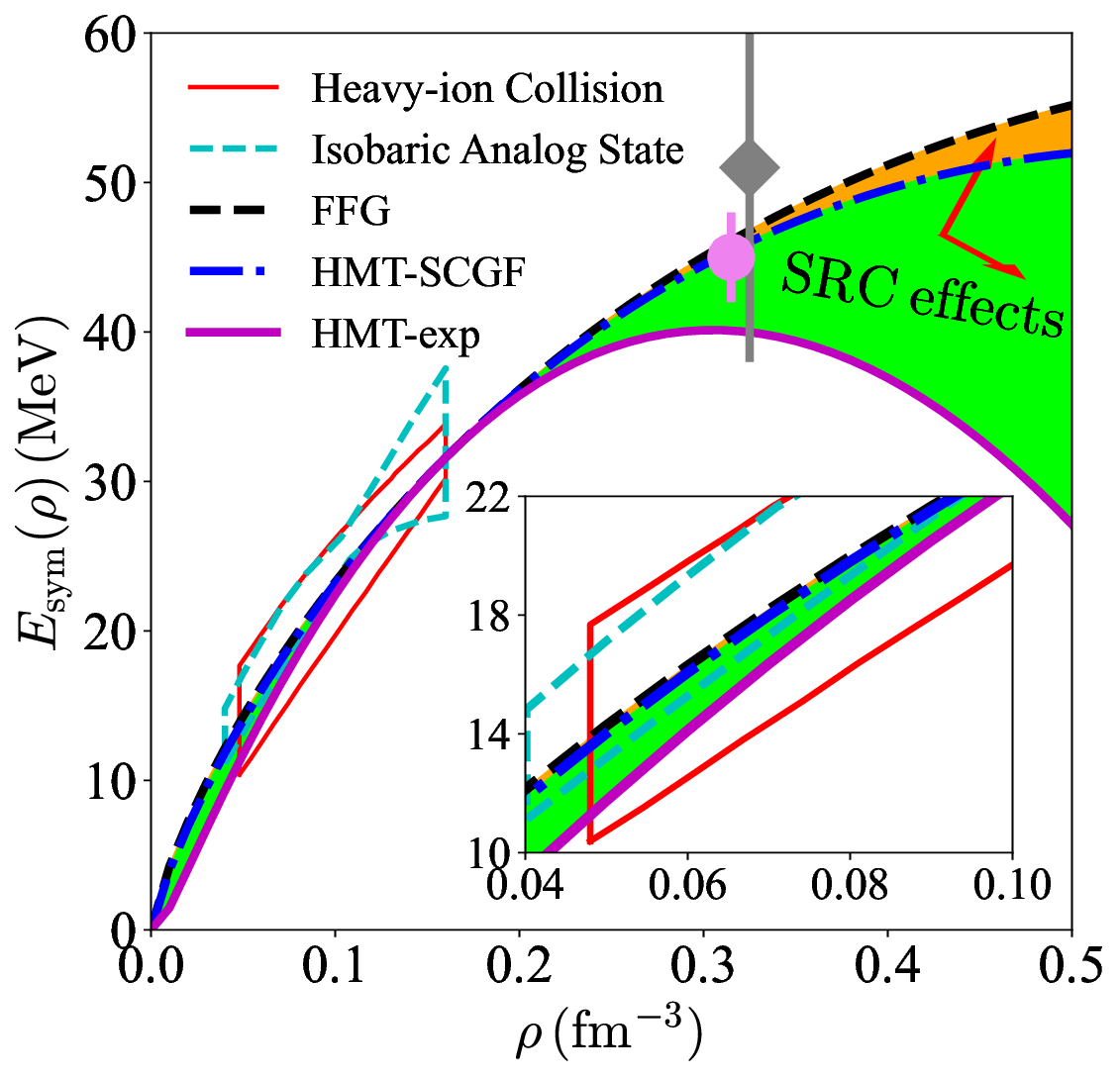}\qquad
  \includegraphics[width=8.5cm]{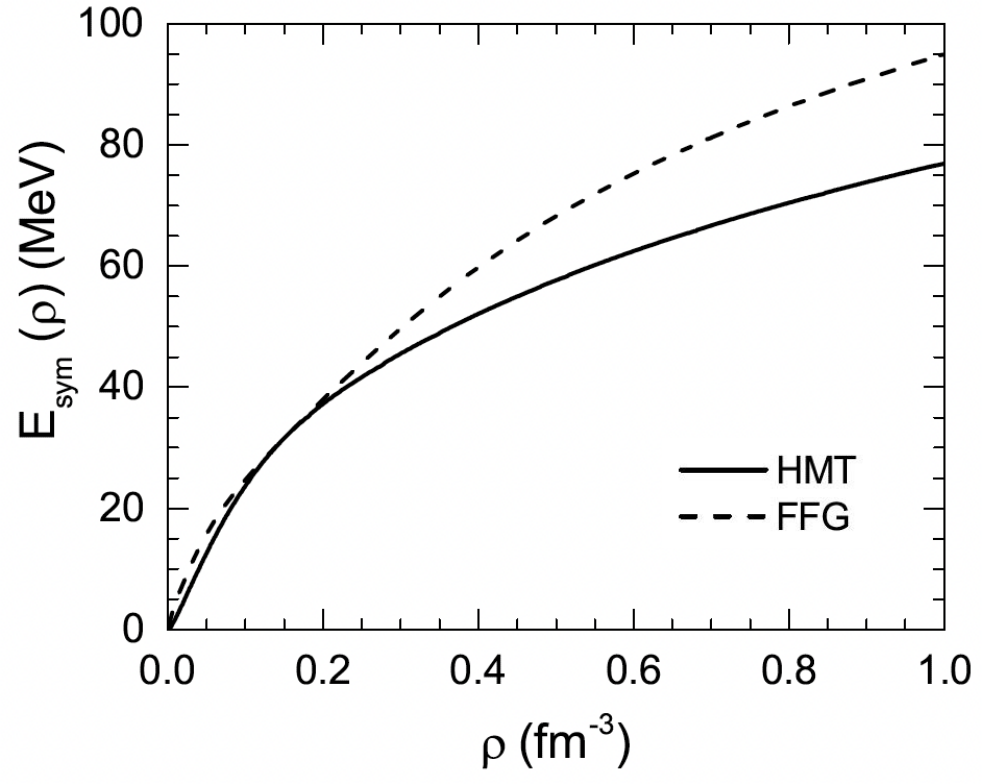} 
  \caption{(Color Online). Left: density dependence of the symmetry energy in the FFG model and HMT model using the Gogny-type momentum-dependent interaction. Constraints from heavy-ion collisions\cite{Tsa12} and the isobaric analog state studies\cite{Dan14} are shown; figure taken from Ref.\cite{CaiLi22Gog}.
  Right: the same as the left panel but using the nonlinear Walecka model; figure taken from Ref.\cite{Cai16b}.
  }
  \label{fig_ab_Esym}
\end{figure}

\begin{figure}[h!]
\centering
 \includegraphics[width=8cm]{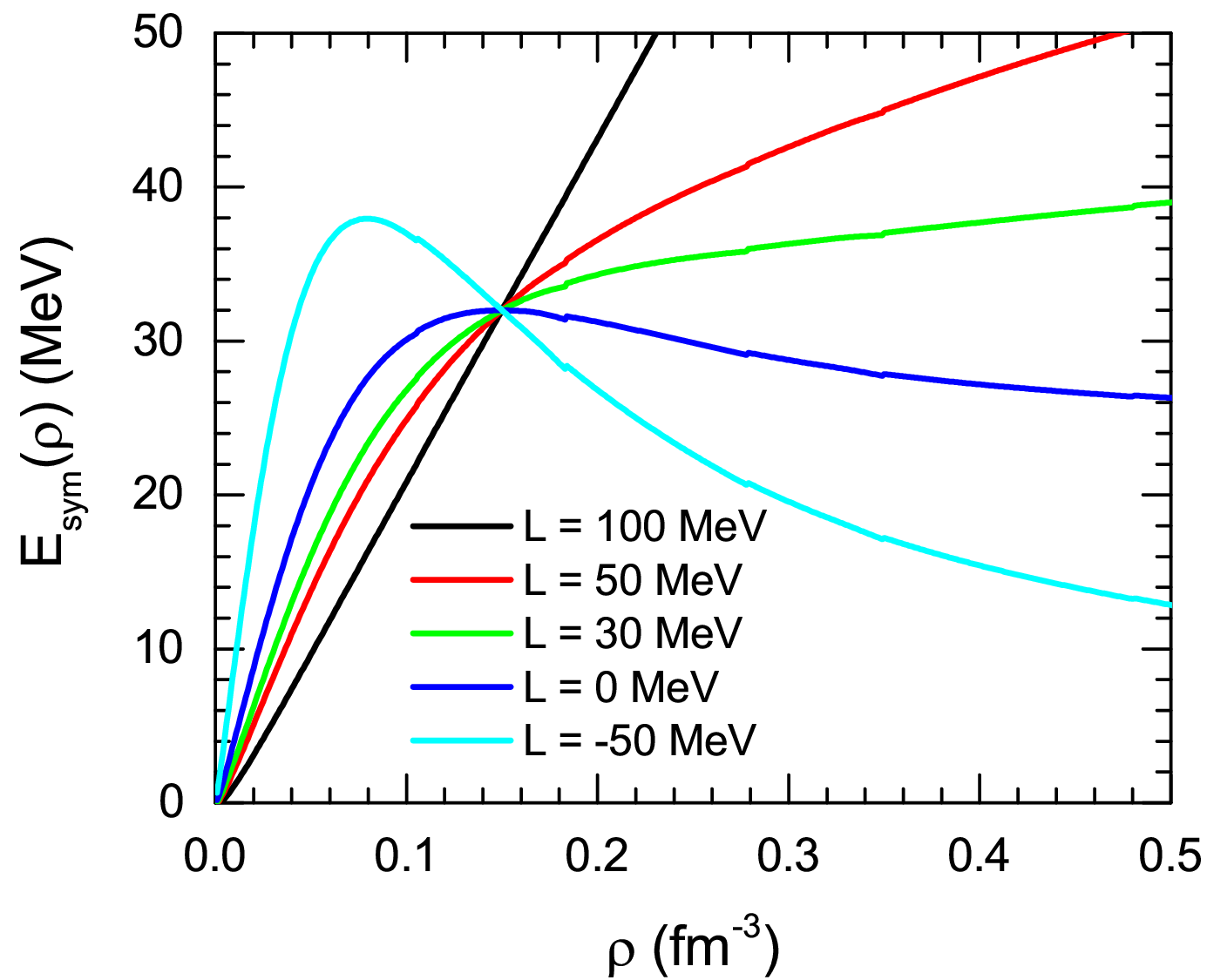}\qquad
 \includegraphics[width=8.5cm]{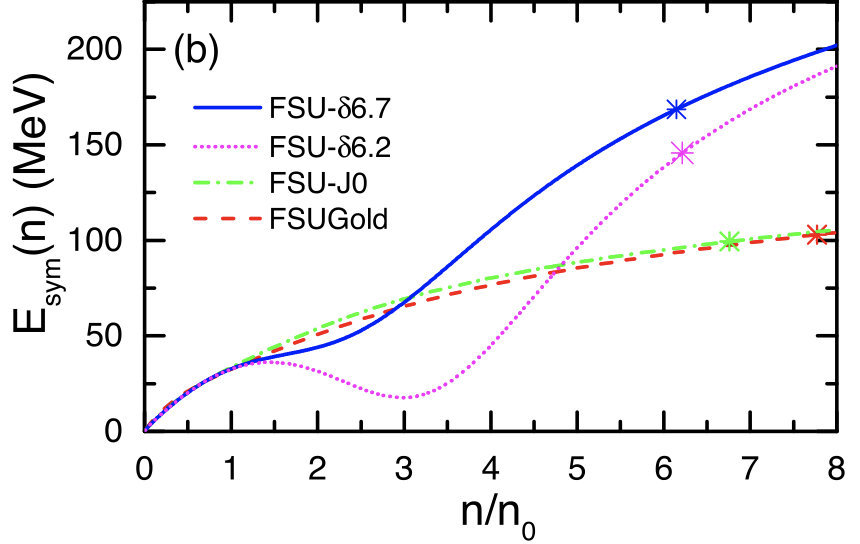}
 \caption{(Color Online). Left: the symmetry energy with different $L$ as a function of density in the nonlinear Walecka model with SRC-induced HMT, the symmetry energy at the saturation density is fixed; figure adapted from Ref.\cite{LCCX18}. Right: the symmetry energy from the nonlinear Walecka model by including the $\sigma$-$\delta$ coupling; figure taken from Ref.\cite{LiF22}.}
  \label{fig-EsymRMFHMT}
\end{figure}

The inclusion of HMT therefore makes $E_{\rm{sym}}^{\rm{kin}}(\rho)$ depend solely on the Dirac effective mass $M_0^{\ast}\equiv M_{\rm N}-g_\sigma\overline{\sigma}$, and leads to a consistently reduced kinetic symmetry energy compared with the FFG prediction. At $M_0^{\ast}/M_{\rm N}=0.6$, for example, one finds $E_{\rm{sym}}^{\rm{kin}}(\rho_0)\approx -16.94\pm13.66\,\rm{MeV}$\cite{Cai16b}, in close agreement with the corresponding non-relativistic estimates (see the left panel of FIG.\,\ref{fig_esym0}). This reduction is thus a robust and model-independent feature of incorporating SRC-induced high-momentum correlations.
The density dependence of the symmetry energy obtained from both the non-relativistic potential model and the relativistic Walecka framework is displayed in FIG.\,\ref{fig_ab_Esym}. As seen in the figure, SRC-induced HMT effects reduce the symmetry energy in both approaches across sub-saturation and supra-saturation densities. In other words, the symmetry energy is softened at high densities while being hardened at low densities.
The corresponding SRC-HMT modifications to the symmetric pressure $P_0(\rho)$ are also examined in both models. Although the qualitative trend is consistent, the magnitude of the change in $P_0(\rho)$ is model-dependent. Such variations can affect NS properties, particularly the mass-radius relation. We shall discuss these implications further in Section \ref{SEC_NS}.

The softening of the symmetry energy at $\rho\gtrsim\rho_0$ induced by the SRC-HMT effects in the nonlinear Walecka model also allows one to generate a flat, or even bending-down, behavior of $E_{\rm{sym}}(\rho)$ at intermediate densities. Such a feature is difficult to achieve in conventional RMF models. In other words, by varying the slope parameter $L$ while keeping $E_{\rm{sym}}(\rho_0)$ fixed, it becomes possible to construct a very soft symmetry energy that even decreases at high densities; see the left panel of FIG.\,\ref{fig-EsymRMFHMT}.
There is circumstantial evidence suggesting that a super-soft $E_{\rm{sym}}(\rho)$ may be required to reproduce the $\pi^-/\pi^+$ ratio in heavy-ion collisions\cite{Xiao09}. Moreover, recent studies incorporating NS astrophysical constraints (e.g., tidal deformabilities) also support this possibility\cite{LiF22,YeJT25,WangSP24,WangSP25}. Illustrated in the right panel of FIG.\,\ref{fig-EsymRMFHMT} are the symmetry energies obtained in a recent RMF model study\cite{LiF22}; in this work, the RMF model can be made simultaneously consistent with: (1) the EOS of symmetric matter at supra-saturation densities extracted from flow data\cite{Dan02} in heavy-ion collisions; (2) the neutron skin thickness of $^{208}$Pb from the PREX-II experiment\cite{PREX}; (3) the largest observed NS mass, from PSR J0740+6620\cite{Riley21,Miller21}; (4) the upper limit $\Lambda_{1.4}\lesssim580$ on the tidal deformability of a $1.4\,M_\odot$ NS from GW170817\cite{Abbott2017}; and (5) the M-R radius measurements of PSR J0030+0451\cite{Riley19,Miller19} and PSR J0740+6620 by NICER\cite{Riley21,Miller21}.
It effectively resolves the well-known tension between PREX-II and GW170817 present in conventional RMF descriptions; as shown in the figure, achieving such consistency requires a flat symmetry energy roughly around $\rho/\rho_0\approx2\sim3$, as exemplified by the FSU-$\delta$6.7 and FSU-$\delta$6.2 parameter sets.
It is also interesting to note here that such a flattened symmetry energy at intermediate densities is closely related to the appearance of a characteristically peaked density profile of the speed of sound in NSs as recently demonstrated quantitatively in Refs.\cite{Zhang:2022sep,Ye:2024meg}.

In short, SRC-induced HMTs impart substantial modifications to the EOS of ANM. Specifically, they stiffen the kinetic EOS of SNM, markedly reduce $E_{\rm{sym}}^{\rm{kin}}(\rho)$, and enhance $E_{\rm{sym,4}}^{\rm{kin}}(\rho)$. To ensure that the total symmetry energy $E_{\rm{sym}}(\rho)$ remains consistent with empirical constraints at $\rho_0$, the reduced kinetic contribution necessitates a corresponding adjustment of the potential part, thereby altering the density dependence of both the isoscalar and isovector channels. In practice, the SRC-modified momentum distribution $n_{\v k}^J$ constitutes the fundamental input for evaluating all kinetic and interaction-related quantities in both non-relativistic and relativistic field frameworks. Across models, the HMT consistently softens $E_{\rm{sym}}(\rho)$ at supra-saturation densities ($\rho \gtrsim \rho_0$) while hardening it at sub-saturation densities, also influencing higher-order isospin-dependent terms and momentum-dependent effects. Moreover, recent studies have extended these analyses to general space dimension $d$\cite{Cai22-dD}, leveraging concepts from ultracold atomic physics\cite{Gio08RMP,Blo08RMP,BECBOOK} and theoretical techniques such as the $\epsilon$-expansion from statistical physics\cite{Wilson1974,Nishida2006,Nishida2011}. These SRC-HMT-induced EOS modifications naturally propagate to the NS structure, as discussed in the following section.

\section{SRC-HMT Effects on Neutron Star Properties: Selected Results with Open Issues}\label{SEC_NS}

\indent 

The SRC-induced HMT, particularly its isospin dependence, is expected to influence the properties of cold NSs, as the neutron-rich environment enhances the proton high-momentum component, thereby amplifying its dynamical impact. The NS mass is determined by integrating the Tolmann--Oppenheimer--Volkoff (TOV) equations (setting $c=1$)\cite{TOV39-1,TOV39-2}:
\begin{align}
\frac{\d P}{\d r} &= -\frac{GM\varepsilon}{r^2}
\left(1+\frac{P}{\varepsilon}\right)
\left(1+\frac{4\pi r^3P}{M}\right)
\left(1-\frac{2GM}{r}\right)^{-1},~~
\frac{\d M}{\d r} = 4\pi r^2\varepsilon,
\end{align}
where $M=M(r)$, $P=P(r)$, and $\varepsilon=\varepsilon(r)$ denote the mass, pressure, and energy density at the radial coordinate $r$. The equations are highly nonlinear and are solved numerically by choosing a central pressure and integrating outward to the stellar radius $R$ where $P(R)=0$ for a given EOS $P(\varepsilon)$. The resulting gravitational mass is $
M_{\rm{NS}} \equiv M(R) = \int_0^R 4\pi r^2 \varepsilon(r)\d r$.

\begin{figure}[h!]
\centering
  \includegraphics[width=17.cm]{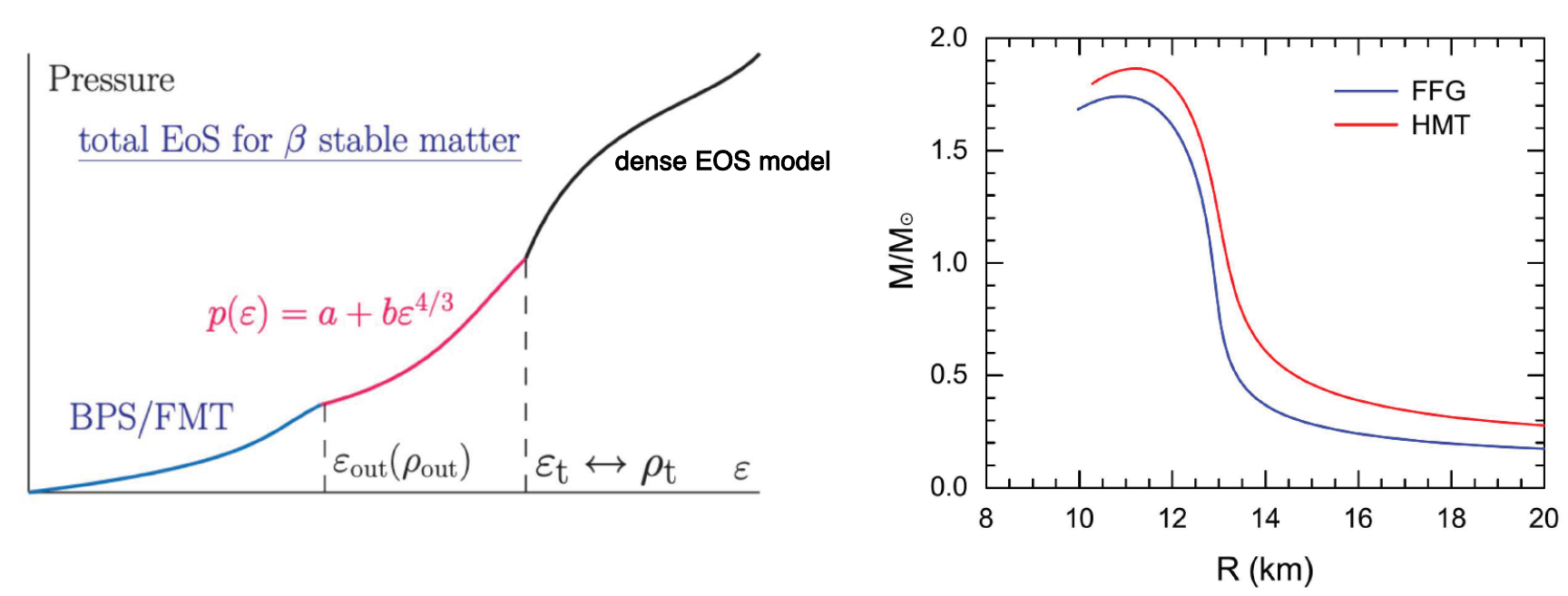}
  \caption{(Color Online). Left: construction of $\beta$-stable EOS in NS calculations. Right: the NS M-R relation obtained by integrating the TOV equation under the EOS of NS matter in the FFG and HMT models, respectively. Figure taken from Ref.\cite{Cai16c}.}
  \label{fig_betaEOS}
\end{figure}

\begin{figure}[h!]
\centering
  \includegraphics[width=17.5cm]{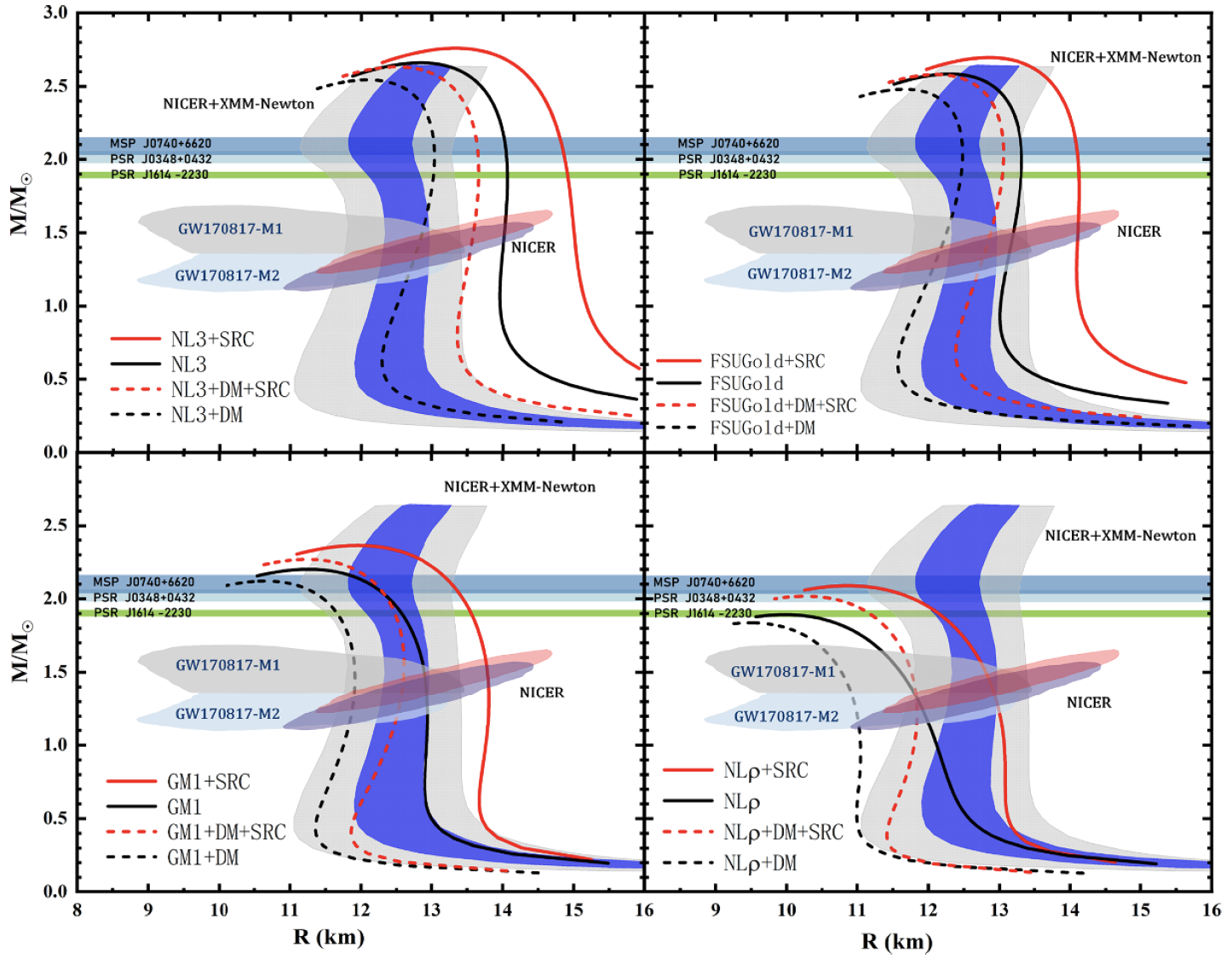}
  \caption{(Color Online). NS M-R relations calculated using four widely employed relativistic parameter sets, each shown in four variants: the original version (black solid), the SRC-revised version (red solid), the dark-matter (DM) revised version (black dashed), and the SRC+DM revised version (red dashed). Figure taken from Ref.\cite{Hong23CQG}.  }
  \label{fig_HongMR}
\end{figure}

The left panel of FIG.\,\ref{fig_betaEOS} illustrates the construction of the $\beta$-equilibrium EOS employed in NS calculations. 
For neutrino-free $\beta$-stable nuclear matter, chemical equilibrium for the processes $\rm{n}\rightarrow\rm{p}+\rm{e}^{-}+\overline{\nu}_{\rm{e}}$ and 
$\rm{p}+\rm{e}^{-}\rightarrow\rm{n}+\nu_{\rm{e}}$ 
requires $ \mu_{\rm{e}}=\mu_{\rm{n}}-\mu_{\rm{p}}
\approx4\delta E_{\mathrm{sym}}(\rho)
+8\delta^{3}E_{\mathrm{sym},4}(\rho)
$, where $\mu_i\,(i=\rm{n,p,e},\mu)$ denotes the corresponding chemical potential. For relativistic degenerate electrons, one has 
$
\mu_{\rm{e}}
=(m_{\rm e}^{2}+k_{\rm F}^{\rm e,2})^{1/2}
=\left[m_{\rm e}^{2}+(3\pi^{2}\rho x_{\rm e})^{2/3}\right]^{1/2}
\approx(3\pi^{2}\rho x_{\rm e})^{1/3}
$, where $m_{\rm e}\approx0.511\,\mathrm{MeV}$ is the electron mass and charge neutrality imposes $x_{\rm p}=x_{\rm e}$.
Once the electron chemical potential exceeds the muon rest mass 
$m_{\mu}\approx0.105\,\mathrm{GeV}$, the reactions 
$\rm{e}^{-}\rightarrow\mu^{-}+\nu_{\rm e}+\overline{\nu}_{\mu}$, 
$\rm{p}+\mu^{-}\rightarrow\rm{n}+\nu_{\mu}$, and 
$\rm{n}\rightarrow\rm{p}+\mu^{-}+\overline{\nu}_{\mu}$ 
become energetically allowed. The appearance of muons modifies the $\beta$-equilibrium conditions to $
\mu_{\rm n}-\mu_{\rm p}=\mu_{\rm e}$ and $
\mu_{\rm n}-\mu_{\rm p}=\mu_{\mu}
=[m_{\mu}^{2}+(3\pi^{2}\rho x_{\mu})^{2/3}]^{1/2},
$ with $x_{\rm p}=x_{\rm e}+x_{\mu}$.
For npe$\mu$ matter, the total energy density is 
$ \varepsilon(\rho,\delta)
=\varepsilon_{\rm N}(\rho,\delta)
+\varepsilon_{\rm e}(\rho,\delta)
+\varepsilon_{\mu}(\rho,\delta),
$ where the nucleonic contribution is that of np matter,
$ \varepsilon_{\rm N}(\rho,\delta)
=[M_{\rm N}+E(\rho,\delta)]\rho,
$ with $E(\rho,\delta)$ the EOS of ANM.
The leptonic energy densities are computed in the FFG approximation, $
\varepsilon_{\ell}(\rho,\delta)
=\eta_{\ell}\phi_{\ell}(t_{\ell})$, with $
\eta_{\ell}
={m_{\ell}}/{8\pi^{2}\lambda_{\ell}^{3}}$, $
\lambda_{\ell}
={m_{\ell}^{-1}}$, $
t_{\ell}
=\lambda_{\ell}(3\pi^{2}\rho_{\ell})^{1/3}$, and 
\begin{equation}
\phi_{\ell}(t_{\ell})
=t_{\ell}\left(1+2t_{\ell}^{2}\right)\sqrt{1+t_{\ell}^{2}}
-\ln\!\left(t_{\ell}+\sqrt{1+t_{\ell}^{2}}\right),
\end{equation}
where $m_{\ell}$ and $\rho_{\ell}$ denote the lepton masses and number densities. The total pressure is then $
P(\rho,\delta)
=P_{\rm N}(\rho,\delta)
+P_{\rm e}(\rho,\delta)
+P_{\mu}(\rho,\delta),
$ with the nucleonic pressure given by the Hugenholtz--Van Hove (HVH) theorem\cite{Hug58,Cai12PLB,XuC2011}, $ P_{\rm N}(\rho,\delta)
=\rho_{\rm p}\mu_{\rm p}+\rho_{\rm n}\mu_{\rm n} -\varepsilon_{\rm N}(\rho,\delta)$. The leptonic contributions satisfy $P_{\ell}(\rho,\delta) =\rho_{\ell}\mu_{\ell} -\varepsilon_{\ell}(\rho,\delta)$, where $\mu_{\ell} =\sqrt{k_{\ell}^{2}+m_{\ell}^{2}}$ and $ k_{\ell}=(3\pi^{2}\rho_{\ell})^{1/3}
$. Within this formalism, the HVH relation, $
P=\rho^{2}\d(\varepsilon/\rho)/\d\rho$ is automatically fulfilled.
The transition density $\rho_{\rm{t}}$ denotes the baryon number density separating the liquid core from the inner crust in NSs\cite{XuJ09PRC}. A simple and widely adopted approach for determining $\rho_{\rm{t}}$ is the thermodynamical method, which requires the system to satisfy the intrinsic stability conditions
\begin{equation}
-\left( \frac{\partial P}{\partial v}\right) _{\mu_{\rm{np}}} > 0,\qquad
-\left( \frac{\partial \mu_{\rm{np}}}{\partial q_{\rm{c}}}\right) _{v} > 0,
\label{ther2}
\end{equation}
where $P$ is the total pressure, $v$ and $q_{\rm{c}}$ denote the volume and charge per baryon, and $\mu_{\rm{np}}=\mu_{\rm{n}}-\mu_{\rm{p}}$ is the neutron-proton chemical potential difference. Using the relations $\partial E(\rho,x_{\rm{p}})/\partial x_{\rm{p}}=-\mu_{\rm{np}}$ and $x_{\rm{p}}=\rho_{\rm{p}}/\rho$, and treating the electrons as a free Fermi gas, one finds that the thermodynamical stability criteria in Eq.\,(\ref{ther2}) are equivalent to requiring the incompressibility of NS matter at $\beta$-equilibrium to stay positive\cite{Lat07,Kubis07,XuJ09PRC}
\begin{align}
K_{\rm{\mu}}(\rho)
=&\;2\rho\frac{\partial E(\rho,x_{\rm{p}})}{\partial \rho}
+\rho^{2}\frac{\partial^{2}E(\rho,x_{\rm{p}})}{\partial \rho^{2}}
-\left.
\left( \frac{\partial^{2}E(\rho,x_{\rm{p}})}{\partial \rho\partial x_{\rm{p}}}
\rho \right)^{2}
\right/
\frac{\partial^{2}E(\rho,x_{\rm{p}})}{\partial x_{\rm{p}}^{2}}
>0.
\label{Vther}
\end{align}
The baryon number density at which Eq.\,(\ref{Vther}) is first violated then defines the core-crust transition density $\rho_{\rm{t}}$; consequently $P_{\rm t}=P(\rho_{\rm t})$.
Moreover, near the very surface of the NS, the EOS is often taken as the BPS/FMT form\cite{BPS71,Iida97}. As the density increases, one encounters the outer crust and eventually reaches the core-crust interface, the $\rho_{\rm t}$ of which is obtained self-consistently within the employed model. For the EOS between the outer crust and the transition boundary, a frequently used parametrization is
\begin{equation}\label{def-inteEOS}
P(\varepsilon)=a+b\varepsilon^{4/3},
\end{equation}
where the coefficients $a$ and $b$ are determined consistently with the calculated transition density. The outer-crust density is taken to be $\rho_{\rm{out}}\approx2.46\times10^{-4}\,\rm{fm}^{-3}$. In the region $6.93\times10^{-13}\,\rm{fm}^{-3}\lesssim\rho\lesssim\rho_{\rm{out}}$, the BPS EOS is employed, while for $4.73\times10^{-15}\,\rm{fm}^{-3}\lesssim\rho\lesssim6.93\times10^{-13}\,\rm{fm}^{-3}$ the FMT EOS is used\cite{BPS71,Iida97}.

\begin{figure}[h!]
\centering
  \includegraphics[height=6cm]{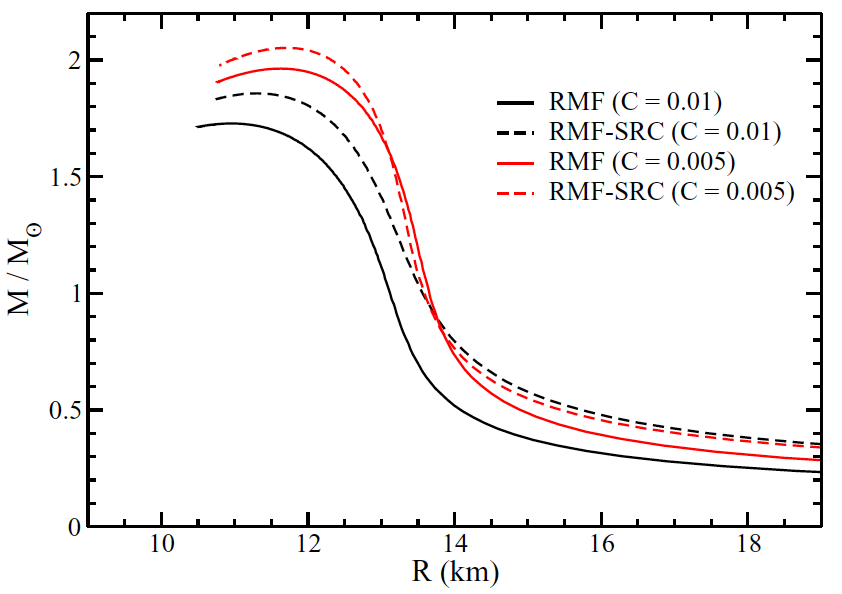}\quad
  \includegraphics[height=6cm]{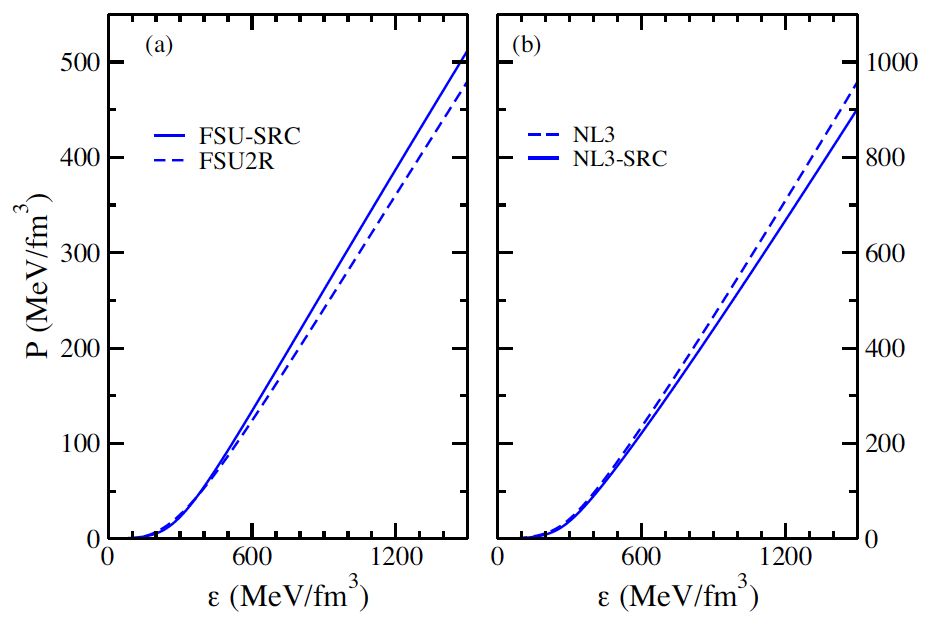}
  \caption{(Color Online). Left: the NS M-R relations in the nonlinear Walecka model with and without SRC-HMT effects. Here, $C$ denotes the strength of the $\omega$-meson self-interaction term, $\sim C(\omega_\mu\omega^\mu)^2$; figure adapted from Ref.\cite{Sou20PRC}. Right: similar to the left panel, but for two parameter sets: one including the $\omega$-meson self-interaction term (FSU-2R)\cite{FSU2R} and one without (NL3)\cite{NL3}; figure adapted from Ref.\cite{Rod23}.
} \label{fig_SouMR}
\end{figure}

SRC-induced HMT effects influence various NS properties, such as the M-R relation and tidal deformabilities, in several ways: (a) they modify the proton fraction $x_{\rm p}$, which directly affects the composition of NS matter and plays a crucial role in determining the NS cooling behavior\cite{Yak01,Dong16ApJ,Dong24EPJA,ShangXL20,LiuCX24}; and
(b) they also reshape the dense-matter EOS at supra-saturation densities, as discussed in the previous section.
Together, these effects can impact the NS mass and radius\cite{Sou20xx,Hong25PLB,Hong23CQG,Gaut25}, tidal deformabilities\cite{Lour22PRD,Lour25xx,Sou20PRC}, oscillation spectra\cite{Hong22CPC,Hong23CQG}, and other macroscopic observables.
Shown in the right panel of FIG.\,\ref{fig_betaEOS} is the predicted NS M-R relation obtained using the FFG and HMT models within the nonlinear Walecka framework\cite{Cai16c}. Although the symmetry energy is softened at high densities, the EOS of SNM is substantially stiffened, leading to an overall enhancement of the maximum NS mass. In particular, the SRC-HMT increases the skewness parameter $J_0$ of the SNM EOS\cite{Cai16c}, which plays a crucial role in determining NS masses\cite{Cai17NST}. In contrast, the impact of the symmetry energy on the M-R relation is comparatively weak in RMF models\cite{Serot1986}. As a result, the increased skewness induced by the HMT elevates the maximum NS mass. Quantitatively, the maximum masses for the HMT and FFG cases are
$M_{\rm{NS}}^{\max}\approx 1.87\,M_\odot$ and
$M_{\rm{NS}}^{\max}\approx 1.74\,M_\odot$,
with corresponding radii of approximately $11.21$\,km and $10.89$\,km\cite{Cai16c}, respectively. This represents an $\sim8\%$ increase in $M_{\rm{NS}}^{\max}$. While the HMT EOS still underpredicts the observed maximum masses, it moves the theoretical predictions closer to current astrophysical measurements\cite{Riley19,Miller19,Riley21,Miller21,Fon21,Choud24,Reard24,Ditt24,Salmi22,Salmi24,Vin24}.

After the work of Ref.\cite{Cai16c}, several related studies have been conducted\cite{Sou20xx,Sou20PRC,Lour22PRD,Lour25xx,Rod23,Hong23CQG,Hong22CPC,Hong25PLB,LuH22NPA}. For instance, Ref.\cite{Sou20xx} employed the same nonlinear Walecka model to examine the NS cooling pattern with and without SRC-HMT effects, and also analyzed the NS M-R relation. Similarly, Ref.\cite{Hong23CQG} adopted a slightly modified parametrization of $n(k)$ in the nonlinear Walecka model, with and without dark-matter (DM) effects, and the resulting M-R relations are shown in FIG.\ref{fig_HongMR}; an enhanced NS maximum mass due to the SRC-HMT is observed. The same conclusion was reported in Ref.\cite{Sou20PRC}, as illustrated in the left panel of FIG.\,\ref{fig_SouMR}.
Subsequently, the role of the $\omega$-meson self-interaction term, $\sim C(\omega_\mu\omega^\mu)^2\sim c_\omega(\omega_\mu\omega^\mu)^2$ (see Eq.\,(\ref{rmf_lag})), was investigated in detail. For example, Ref.\cite{Rod23} found that including this self-interaction term leads the SRC-HMT to enhance the maximum mass (as shown in the left panel of FIG.\,\ref{fig_SouMR}), whereas in its absence the SRC-HMT tends to reduce the maximum mass, see the right panel of FIG.\,\ref{fig_SouMR} where two parameter sets are constructed: one including the $\omega$-meson self-interaction term (FSU-2R)\cite{FSU2R} and one without (NL3)\cite{NL3}.
Thus, there is no definitive evidence as to whether SRC-HMT universally increases or decreases $M_{\rm{NS}}^{\max}$. A similar trend was observed in the modified Gogny-type model with a momentum-dependent potential\cite{CaiLi22Gog}, where incorporating SRC-HMT effectively reduces $M_{\rm{NS}}^{\max}$.
More related works on NSs with DM effects can be found in Refs.\cite{Pan17,John18,Nelson19,Das21,Liu23DM,Zhang25DM,Sun24DM,Raj18,Gara19,Bell20,Bell21,Liu25DM,Hipp23DM,Kumar24,Maha25}.
Furthermore, Ref.\cite{Hen16-gamma} emphasized that although NS observables are sensitive to the high-density behavior of the symmetry energy, they do not provide sufficient discriminatory power to distinguish between HMT and FFG descriptions of the kinetic contribution. Their analysis suggests that the NS EOS inferred from Bayesian analyses of NS observations is robust and largely insensitive to the specific microscopic modeling of $E_{\rm{sym}}^{\rm{kin}}$. Such insensitivity, or the ``blindness'' of the TOV equations, is consistent with recent findings based on a dimensionless reformulation of the TOV equations\cite{CaiLi25-IPADTOV}.

\begin{figure}[h!]
\centering
\includegraphics[width=8.5cm]{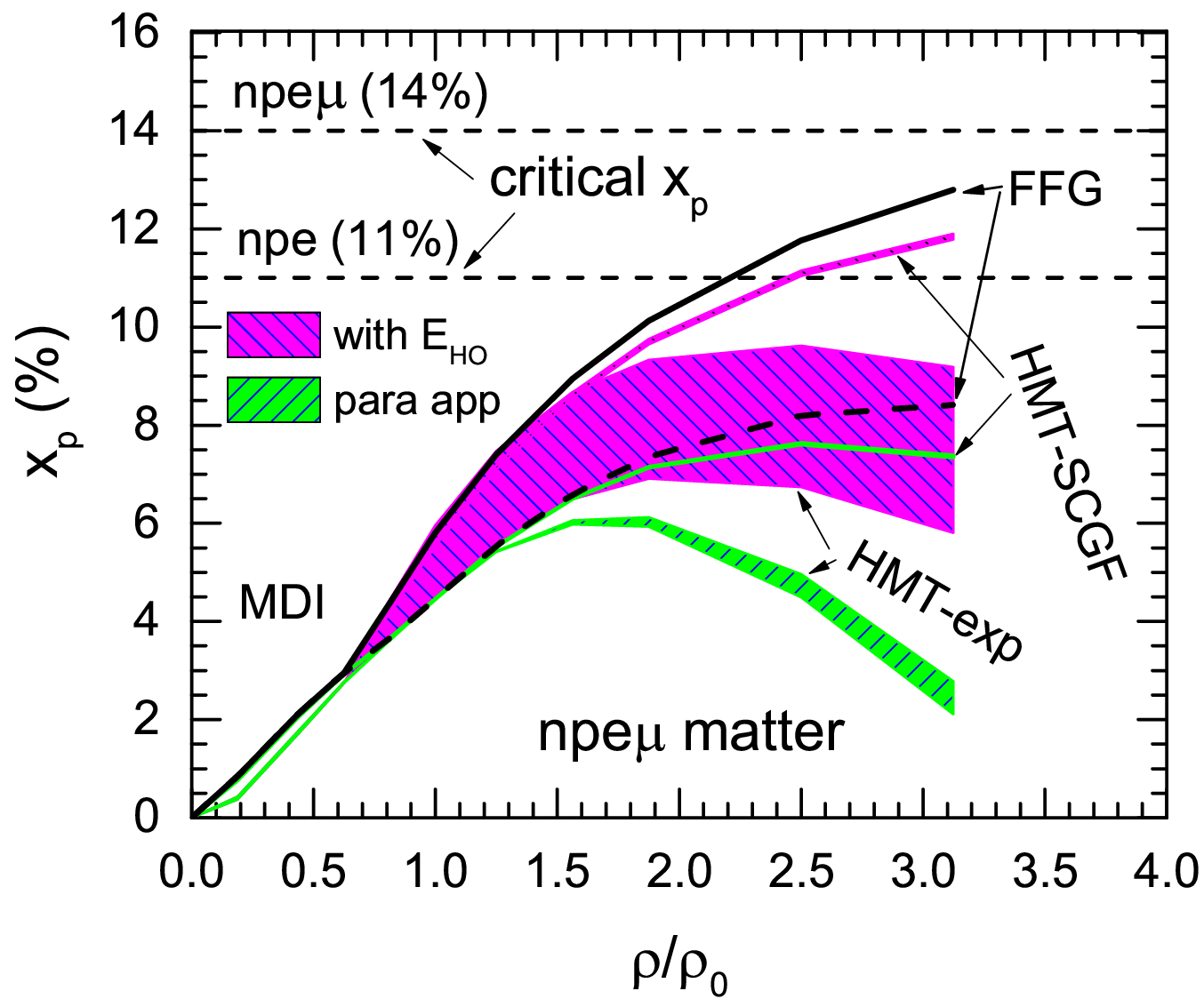}
\caption{(Color Online). Proton fraction $x_{\rm{p}}$ obtained in the modified Gogny-type model with/without the SRC-induced HMT.
The shadow regions are caused by the uncertainties of the macroscopic quantities of the ANM EOS.}
\label{fig_ab_xp-1}
\end{figure}

\begin{figure}[h!]
\centering
  \includegraphics[width=17.5cm]{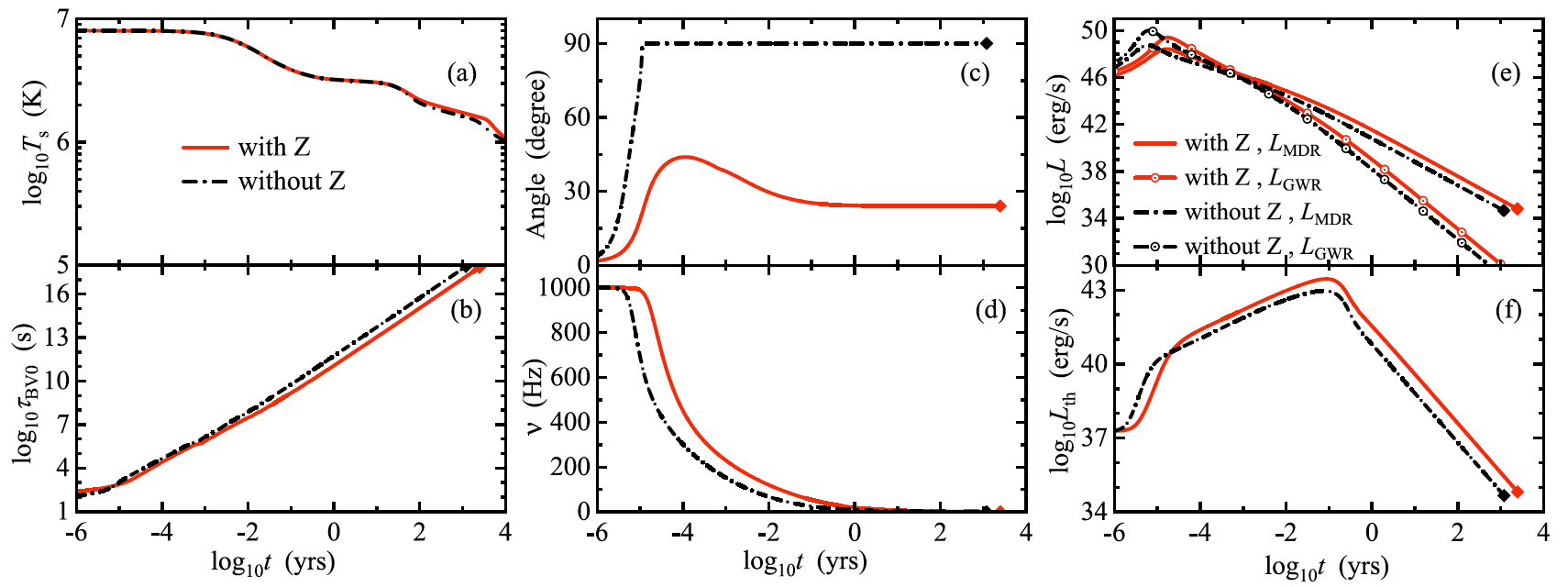}
  \caption{(Color Online). Impact of the Migdal--Luttinger $Z$-factor on the evolution of a canonical magnetar ($1.4\,M_\odot$, $P_0=1$\,ms), showing its effects on surface temperature $T_{\rm s}$, spin frequency $\nu$, magnetic inclination $\chi$, damping timescale $\tau_{\rm{BV0}}$, and luminosities $L_{\rm{GW}}$, $L_{\rm{MDR}}$, and $L_{\rm{th}}$; the diamond marks when $T$ reaches the superfluid critical temperature $T_{\rm c}$. Figure taken from Ref.\cite{LiuCX24}.}
  \label{fig_LiuNS}
\end{figure}

As discussed above, the SRC-induced HMT can significantly influence the proton fraction $x_{\rm p}$, which is crucial for determining the NS cooling pattern\cite{Yak01,Fra08}.  
In the simplest approximation, $x_{\rm p}$ can be estimated from the symmetry energy as
\begin{equation}\label{def_xp_z1}
\boxed{
x_{\rm{p}}\approx\frac{1}{2}\left/\left[3+\left(\frac{k_{\rm{F}}}{4E_{\rm{sym}}(\rho)}\right)^3\right]\right..}
\end{equation}
This expression indicates that at very low densities, $k_{\rm{F}}/4E_{\rm{sym}}(\rho)\to 3M_{\rm N}/2k_{\rm{F}}\sim\rho^{-1/3}$, so that $x_{\rm{p}}\sim \rho$ in this limit.  
In the high-density limit, for instance, in the nonlinear Walecka model (of (\ref{rmf_lag})) using a FFG description, the symmetry energy leads to
\begin{equation}
x_{\rm{p}}\to\frac{1}{2}\left/\left[3+\left[\left(\frac{3\pi^2}{2}\right)^{1/3}\left/\left[\frac{2}{3}\left(\frac{3\pi^2}{2}\right)^{1/3}+\frac{2c_{\omega}^{2/3}}{\Lambda_{\rm{V}}}\right]\right.\right]^3\right]\right.,
\end{equation}
with a typical value around 10\%, depending on the magnitudes of $c_\omega$ and $\Lambda_{\rm V}$\cite{CaiLi22PRCFFG}, e.g., $x_{\rm p}\to4/51\approx7.8\%$ if $c_\omega=0$, or $x_{\rm p}\approx15\%$ for $c_\omega^{2/3}/\Lambda_{\rm V}\approx1$; $x_{\rm p}\lesssim1/6\approx16\%$ can be obtained for $c_{\omega}^{2/3}/\Lambda_{\rm V}\to\infty$.
Due to the complexities of higher-order terms in the EOS of ANM when the HMT is included, we approximate these contributions when determining the $x_{\rm p}$ as
$8E_{\rm{sym,4}}(\rho)\delta^3+\cdots\approx 8E_{\rm{HO}}(\rho)\delta^3$ with
$E_{\rm{HO}}(\rho)=E_{\rm{sym,4}}(\rho)+E_{\rm{sym,6}}(\rho)+\cdots=E_{\rm{n}}(\rho)-E_0(\rho)-E_{\rm{sym}}(\rho)$, where $E_{\rm{n}}(\rho)$ is the EOS of PNM. As shown in Ref.\cite{Cai12PRC-S4}, including the fourth-order symmetry energy yields $x_{\rm p}$ values very close to the exact result. This scheme is partly necessitated by the difficulty that the chemical potential is not generally given by $k_{\rm{F}}^{\rm{n/p},2}/2M_{\rm N}+U_{\rm{n/p}}(\rho,\delta,k_{\rm{F}}^{\rm{n/p}})$ once the HMT in $n_{\v{k}}^J$ is considered\cite{AGD}.
FIG.\,\ref{fig_ab_xp-1} displays the density dependence of the proton fraction $x_{\rm p}$ in $\beta$-stable npe$\mu$ matter, obtained within the modified Gogny-type model with and without SRC-HMT effects. The shaded regions account for uncertainties in macroscopic quantities such as $K_0$ and $L$. Three main observations can be made:  
(a) Including the SRC-HMT softens the symmetry energy at supra-saturation densities and reduces $x_{\rm p}$ at fixed densities relative to the FFG prediction (black line vs. thin magenta and green bands);  
(b) as the fraction of high-momentum nucleons in finite nuclei and nuclear matter increases, e.g., from the HMT-SCGF set to the HMT-exp set\cite{Cai16b}, the supra-normal density symmetry energy becomes softer and the corresponding $x_{\rm p}$ is further reduced;  
(c) the fourth-order symmetry energy significantly affects $x_{\rm p}$; e.g., in the HMT-exp set it increases $x_{\rm p}$ at $\rho=3\rho_0$ from approximately 3.5\% to 7.5\%. 

Notably, within the HMT-exp set\cite{Cai16b}, the maximum proton fraction $x_{\rm p}$ never surpasses the threshold for the direct Urca (DU) process, i.e., $\sim 11\%$ for npe matter and $\sim 14\%$ for npe$\mu$ matter.  
This naturally raises the question of whether the DU process is ultimately favorable, considering the SRC-HMT effects. Addressing this issue is highly nontrivial for at least two reasons:  
(a) The HMT above the Fermi surface may effectively open additional channels for the DU process, potentially lowering the effective proton fraction threshold below the nominal 11\% or 14\% values\cite{Fra08,LCCX18};  
(b) Conversely, the combined effect of the HMT and the low-momentum depletion of $n(k)$ significantly reduces the Migdal--Luttinger jump $Z_{\rm F}^J$ at the Fermi surface, which in turn strongly affects the neutrino emissivities\cite{Dong16ApJ,Dong24EPJA}.  
Therefore, a rigorous analysis of NS cooling that includes SRC-HMT effects is inherently complex and typically requires more sophisticated treatments, such as those based on the spectral function approach\cite{Sed24PRL}.
As an illustration, FIG.\,\ref{fig_LiuNS} displays the influence of the Migdal--Luttinger $Z$-factor on various evolutionary properties of a canonical magnetar with mass about $1.4\,M_{\odot}$; the figure shows that the $Z$-factor significantly affects some quantities, such as the magnetic inclination angle, while having minimal impact on others.
There are numerous intriguing aspects of NS cooling, including the existence/disappearance of the crustal phase structures\cite{Peli23EPJA}, mean free path calculations\cite{Neg81,Fang14PRC}, and other transport properties such as thermal and electrical conductivities and viscosities\cite{SXLi11PRC,XuJ13PLB,CLZhou13PRC,LiuC22,Deng24PPNP,Ma23,Deng22PRC,Deng16PRC}. We do not attempt a detailed review of these issues here and refer interested readers to Ref.\cite{CLMRev25} and the references therein.

In short, the SRC-induced HMT, especially its isospin dependence, significantly impacts cold NSs by modifying the proton fraction $x_{\rm p}$ and reshaping the dense-matter EOS at supra-saturation densities, which in turn affects the NS M-R relation, tidal deformabilities, oscillation modes, and other macroscopic observables. While the HMT can increase $M_{\rm{NS}}^{\max}$ by stiffening the SNM EOS, its effect is strongly model-dependent and intertwined with factors like the $\omega$-meson self-interaction and the Migdal--Luttinger $Z$-factor, making its influence on the DU process highly nontrivial. Moreover, NS cooling and transport properties, including crustal phases, mean free paths, and viscosities, further complicate the analysis, highlighting the need for detailed studies such as those based on the spectral function approach.
In the following, we highlight a few open issues to stimulate further studies on SRC-HMT effects and their impact on certain NS properties.

\subsection*{\color{teal}{(1) What is the High-momentum Nucleon Fraction at NS Densities?}}

\indent 

\begin{figure}[h!]
\centering
  \includegraphics[width=10.cm]{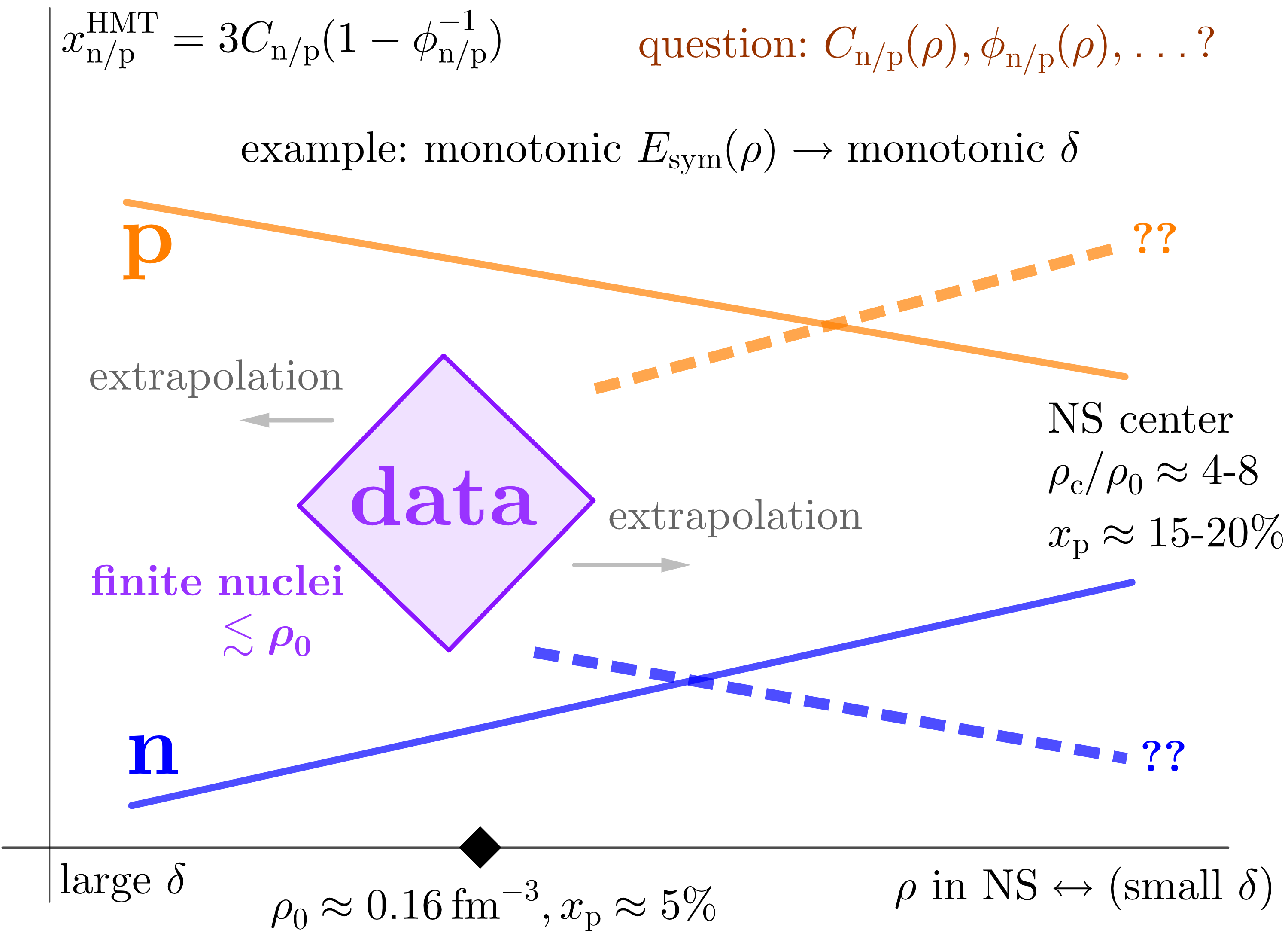}
\caption{(Color Online). A sketch of high-momentum fractions of nucleons as functions of nucleon density. Does the high-momentum nucleon fraction at NS densities exhibit a $\delta$-dependence similar to that observed near saturation density? Available data currently only constrain the behavior at $\rho \lesssim \rho_0 \approx 0.16\,\mathrm{fm}^{-3}$. In this sketch, we assume that the symmetry energy $E_{\rm{sym}}(\rho)$ increases monotonically with density. Considering two connected regions with local densities $\rho_1$ and $\rho_2$, respectively, the isospin-equilibrium requires that\cite{shi} $E_{\rm{sym}}(\rho_1)\delta(\rho_1)=E_{\rm{sym}}(\rho_2)\delta(\rho_2)$. Consequently, the isospin asymmetry $\delta$ in the inner core of a NS is smaller than that near its surface.}\label{fig_xJHMT_NS}
\end{figure}

Up to now, all information about the nucleon SRC-induced HMT is constrained only at densities $\rho\lesssim\rho_0$, either from experimental analyses\cite{Hen14,Sub08} or from microscopic many-body calculations\cite{Rio09,Rio14,Ben93,Benhar08RMP}. However, when studying NSs, these constraints are implicitly extrapolated to much higher densities. For a typical NS, its central density $\rho_{\rm c} \approx 5\rho_0$ corresponds to a characteristic Fermi momentum
$
5^{1/3} k_{\rm F}(\rho_0) \approx 450\,\mathrm{MeV},
$
which lies well within the momentum range dominated by SRCs, so SRC is relevant near NS centers.
A natural question then arises: how reliable is this extrapolation? For instance, in finite nuclei or nuclear matter near saturation density, the high-momentum nucleon fractions exhibit a clear isospin dependence for neutrons and protons, as shown in the right panel of FIG.\,\ref{fig_a2exp} and sketched by the solid lines in FIG.\,\ref{fig_xJHMT_NS}. Yet at supranuclear densities, there is no {\it a priori} reason to expect $x_J^{\rm{HMT}}(\delta)$ to follow the same or even a similar trend; alternative behaviors, such as those suggested by the dashed lines in FIG.\,\ref{fig_xJHMT_NS}, are also plausible.
Because the $\delta$-dependence of $x_J^{\rm{HMT}}$ directly affects the kinetic contributions, the symmetry energy, and the composition of dense matter, uncertainties in its high-density behavior can have significant consequences for predictions of NS properties.

Put it in other words, the HMT-related parameters, such as the contact $C_{\rm{n/p}}$ and the high-momentum cutoff $\phi_{\rm{n/p}}$, are also expected to depend on the density, namely $C_{\rm{n/p}}\to C_{\rm{n/p}}(\rho)$, $\phi_{\rm{n/p}}\to \phi_{\rm{n/p}}(\rho)$.
In this case, the high-momentum fraction $x_J^{\rm{HMT}}$ may exhibit a nontrivial $\delta$-dependence at different densities. To better constrain the HMT parameters at supra-saturation densities, one may develop more advanced theoretical approaches or propose direct experimental probes, e.g., through heavy-ion collision processes\cite{LiBA2008,ZhaoJ24,Tsa12,Xiao09,WX25}.
In Ref.\cite{Yang19PRC}, the momentum distributions at densities beyond the saturation are studied within an extended Brueckner--Hartree--Fock (eBHF) approach, which may provide useful information on $C_{\rm{n/p}}$ at these densities. With improved constraints, one could systematically investigate SRC-HMT effects on NS properties by combining: (a) their impact on the EOS of ANM, particularly the symmetry energy, and (b) their influence on the $\delta$-dependence of the HMT parameters. These two aspects are clearly intertwined, since $E_{\rm{sym}}(\rho)$ directly determines the proton fraction in NSs, and their coupling may lead to nontrivial implications for NS physics.

\subsection*{\color{teal}{(2) Can the SRC-HMT Affect the Inner-crust EOS Index $\gamma$ of $P=a+b\varepsilon^{\gamma}$?}}

\indent

The EOS of the inner crust, spanning the region between the outer crust and the core-crust transition density $\rho_{\rm t}$, is often parametrized using Eq.\,(\ref{def-inteEOS}), i.e., $P \approx a + b \varepsilon^{4/3}$\cite{Carr03}. Since the core-crust transition density, $\rho_{\rm t} \approx 2\rho_0/3$, lies within the density range where SRCs are significant\cite{Hen14}, it is natural to expect that SRCs may also affect the inner crust EOS. In particular, an interesting question is whether the index $\gamma$ in $P \approx a + b \varepsilon^\gamma$ could be modified by SRC-HMT effects. To date, dedicated studies exploring this possibility remain scarce.

The inner crust EOS is further intertwined with exotic nuclear structures, including pasta phases\cite{Sch13,Cap17,Newton22}, cluster formation\cite{WangR25,ZhouB19,XQXu25,DYT25,WYY24,Xu24}, and other effective low-dimensional configurations\cite{Siemens1967,Wong1972,Wong1973,Cao2019}. Its properties are also sensitive to the symmetry energy of dilute nuclear matter\cite{Nat10}. While the inner crust EOS may have a limited influence on the mass and radius of massive NSs, it can significantly affect the radius of low-mass NSs. Moreover, it plays an important role in NS cooling and neutrino emission processes, since an enhanced symmetry energy due to cluster correlations\cite{Nat10} would increase the proton fraction $x_{\rm p}$, thereby modifying related processes\cite{Reddy98PRD,Reddy99PRC}.

Overall, these considerations suggest that a more systematic study of SRC effects on the inner crust EOS, including their interplay with cluster formation, pasta phases, and the density dependence of the symmetry energy, could provide valuable insights into NS structure, thermal evolution, and neutrino dynamics\cite{JY25,Reddy98PRD,Reddy99PRC,Yak01}.

\subsection*{\color{teal}{(3) Does the SRC-HMT Enhance or Reduce the Maximum NS Mass?}}

\indent 

We have discussed that the SRC-HMT may either enhance or reduce the maximum NS mass, depending on the model or the specific nucleon-nucleon interaction employed. Specifically, the effect is sensitive to whether the $\omega$-meson self-interaction term $\sim c_\omega (\omega_\mu \omega^\mu)^2$ is included\cite{Sou20PRC,Lour22PRD,Lour25xx}; similarly, in a Gogny-type model with momentum-dependent interactions, the SRC-HMT has been shown to reduce the maximum mass\cite{CaiLi22Gog}. Moreover, modifying the fitting scheme of the model can change the quantitative impact of SRC-HMT on NS mass\cite{Gaut25}. This naturally raises the question: what is the underlying physical mechanism by which SRC-HMT influences the maximum NS mass?

Some clues emerge from previous studies. It is well known that the effect of the symmetry energy on the maximum NS mass is relatively smaller than that of the symmetric part of the EOS\cite{Serot1979}. In Ref.\cite{Cai16c}, we found that the SRC-HMT induces a larger modification in the EOS of SNM than in the symmetry energy; as a result, the stiffening of the SNM EOS dominates over the softening of the symmetry energy, leading to an increase in the maximum NS mass. In contrast, non-relativistic Gogny-type calculations show that the SRC-HMT has little impact on the SNM EOS, while its effect on the symmetry energy is significant (left panel of FIG.\,\ref{fig_ab_Esym}).
Another relevant factor is the manner in which nuclear parameters are fixed. 
In our studies, the saturation-density parameters are fixed to empirical values: specifically, the SNM EOS is constrained up to order $J_0$, and the symmetry energy up to its slope $L$, before examining the impact of SRC-HMT on higher-order terms, such as the curvature coefficient $K_{\rm{sym}} \equiv 9\rho_0^2\d^2 E_{\rm{sym}}(\rho)/\d\rho^2|_{\rho_0}$ of nuclear symmetry energy.
It is therefore natural that if higher-order nuclear coefficients are also fixed, the SRC-HMT may eventually produce little impact, as the EOSs with and without SRC-HMT would behave similarly up to densities relevant for NS cores.

Overall, systematic investigations along this line are essential to clarify the intrinsic physical mechanisms by which SRC-HMT affects NS mass, radius, and related observables such as tidal deformabilities. Future studies could explore (i) the interplay between SRC-HMT and higher-order EOS parameters, (ii) model-independent constraints on SRC-HMT effects using ab initio calculations or Bayesian analyses of NS observations, and (iii) the combined impact of SRC-HMT on both the SNM EOS and the symmetry energy at supra-saturation densities. Such efforts would provide a more comprehensive understanding of the role of SRCs in dense nuclear matter and their astrophysical implications.

\subsection*{\color{teal}{(4) How Does the SRC-HMT Affect the Formation of Exotics?}}

\indent

Exotic particles, such as hyperons and $\Delta$ resonances, are expected to appear in NS cores; SRC-HMT is expected to influence their formation patterns.
In this context, for example, both the kinetic and potential components of the symmetry energy have been found to separately influence the formation densities of the $\Delta(1232)$ states\cite{Cai15Delta}, using the nonlinear RMF models with a FFG $n(k)$. Specifically, for the reactions 
$\Delta^{++} + \rm{n} \to \rm{p} + \rm{p}$, $\Delta^{+} + \rm{n} \to \rm{n} + \rm{p}$, $\Delta^{0} + \rm{p} \to \rm{p} + \rm{n}$, and $\Delta^{-} + \rm{p} \to \rm{n} + \rm{n}$, 
the critical densities for $\Delta$ formation are determined by\cite{Cai15Delta}:
\begin{align}
\rho_{\Delta^-}^{\rm{crit}}:\;&k_{\rm{F}}^{\rm{n},2}/2M_{\rm{0}}^{\ast}
\approx \Phi_{\Delta}+g_{\sigma\rm{N}}(1-x_{\sigma})\overline{\sigma}-
g_{\omega\rm{N}}(1-x_{\omega})\overline{\omega}_0
-\left[6(1-x_{\rho})E_{\rm{sym}}^{\rm{pot}}(\rho)
+4E_{\rm{sym}}^{\rm{kin}}(\rho)\right]\delta,\\
\rho_{\Delta^0}^{\rm{crit}}:\;&k_{\rm{F}}^{\rm{n},2}/2M_{\rm{0}}^{\ast}
\approx \Phi_{\Delta}+g_{\sigma\rm{N}}(1-x_{\sigma})\overline{\sigma}-
g_{\omega\rm{N}}(1-x_{\omega})\overline{\omega}_0
-2(1-x_{\rho})E_{\rm{sym}}^{\rm{pot}}(\rho)\delta,\\
\rho_{\Delta^+}^{\rm{crit}}:\;&k_{\rm{F}}^{\rm{p},2}/2M_{\rm{0}}^{\ast}
\approx \Phi_{\Delta}+g_{\sigma\rm{N}}(1-x_{\sigma})\overline{\sigma}-
g_{\omega\rm{N}}(1-x_{\omega})\overline{\omega}_0
+2(1-x_{\rho})E_{\rm{sym}}^{\rm{pot}}(\rho)\delta,\\
\rho_{\Delta^{++}}^{\rm{crit}}:\;&k_{\rm{F}}^{\rm{p},2}/2M_{\rm{0}}^{\ast}
\approx \Phi_{\Delta}+g_{\sigma\rm{N}}(1-x_{\sigma})\overline{\sigma}-
g_{\omega\rm{N}}(1-x_{\omega})\overline{\omega}_0
+\left[6(1-x_{\rho})E_{\rm{sym}}^{\rm{pot}}(\rho)
+4E_{\rm{sym}}^{\rm{kin}}(\rho)\right]\delta,
\end{align}
where $\Phi_{\Delta}\approx293\,\rm{MeV}$ is the mass difference between the $\Delta(1232)$ and the nucleon, $M_{\rm{0}}^{\ast}$ is the Dirac nucleon effective mass in SNM, $g_{\sigma\rm{N}}$ and $g_{\omega\rm{N}}$ are the meson-nucleon coupling constants, and $x_{\sigma}, x_{\omega}, x_{\rho}$ characterize the $\Delta$-meson couplings. These relations are obtained within a nonlinear RMF framework that explicitly includes $\Delta$-meson interactions.
This formalism enables a clear separation of the contributions from the kinetic and potential parts of the symmetry energy to the determination of $\rho_{\Delta}^{\rm{crit}}$. In particular, SRCs have been suggested to induce a negative kinetic contribution to the symmetry energy already at normal nuclear density (left panel of FIG.\,\ref{fig_esym0}). Under such conditions, the relative appearance densities of the $\Delta^-$ and $\Delta^{++}$ resonances, as well as the resulting particle fractions and NS structure, may differ from conventional expectations. This suggests that a separate treatment of the kinetic and potential components of the symmetry energy could be relevant for describing certain NS properties\cite{Cai15Delta}.

Future studies could generalize such equations to include SRC-HMT effects on both the kinetic and potential symmetry energy, quantify how large kinetic terms influence the critical densities for $\Delta$ and hyperon formation, and assess the impact on NS composition and dynamical processes. In addition, it is important to explore how SRC-HMT affects in-medium $\Delta$-nucleon and hyperon-nucleon interactions\cite{JHC25}, to develop a self-consistent theoretical framework, and to investigate the connection with the in-medium pion and $\rho$ meson properties\cite{WDF25,Ericson1988,Brown:1990kj}, which could further influence the production and behavior of exotic particles in NS cores.

\subsection*{\color{teal}{(5) Are There Clear Probes of SRC-HMT Effects in NSs?}}

\indent 

The effects of SRC-HMT on NS properties are usually incorporated through modifications of the EOS in the literature. Specifically, studies often examine how observables such as the M-R relations, tidal deformabilities, and proton fractions respond to the inclusion of SRC-induced HMT effects (via the $n(k)$ in the quasi-particle picture). However, it is well known that the TOV equations are, in a sense, composition-blind: given any EOS, they can be integrated to produce M-R relations regardless of the underlying composition of the star. Consequently, the SRC-HMT effects could, in principle, be effectively mimicked by other physical ingredients in the EOS, raising a fundamental question: can we identify intrinsic and unambiguous probes that directly reveal the presence of SRC-HMT effects in NSs?
We notice that similar questions can be raised about all possible new degrees of freedom or phases in NSs.

Identifying intrinsic and differentiable probes for SRC-HMT effects in NSs remains a crucial challenge. Recently, Ref.\cite{Fuku25M} showed that the magnetic-type Love number is a robust indicator for distinguishing NSs from quark stars (QSs), whereas the electric-type Love number may fail when QSs and NSs have similar masses and radii. In a similar spirit, it is necessary to go beyond simply exploring EOS modifications and seek observables that can uniquely and unambiguously reveal SRC-HMT effects, providing a pathway to directly detect and constrain these SRCs in NS interiors.

\section{Summary}\label{SEC_CC}

\indent 

In summary of this brief review, we emphasize the following among many interesting physics in this exciting field:
\begin{itemize}
  \item \textbf{Microscopic origin of SRC--HMTs:} 
  Nucleon SRCs, generated by the tensor and short-range components of the nuclear force, deplete the Fermi sea and produce a characteristic HMT in the single-nucleon momentum distribution, approximately following a universal $k^{-4}$ behavior and dominated by neutron-proton isosinglet pairs.

  \item \textbf{Modification of the kinetic EOS:} 
  SRC--HMTs substantially alter the kinetic contribution to the dense-matter EOS, stiffening the kinetic energy of symmetric nuclear matter while strongly reducing the kinetic symmetry energy and generating a sizable isospin-quartic term in the EOS of ANM.

  \item \textbf{Rearrangement of the potential contribution:} 
  To satisfy empirical constraints near saturation density, the potential part of the EOS must compensate for SRC-induced kinetic effects, leading to significant changes in the density dependence of both isoscalar and isovector interaction channels.

  \item \textbf{Implications for NS properties:} 
  At supranuclear densities relevant to NS interiors, SRC-HMT effects modify the proton fraction, the pressure-energy relation, and TOV solutions, thereby impacting the maximum NS mass, direct URCA cooling thresholds, crust-core transition density, transport properties, critical appearance densities of new particles and phases, and tidal responses in a currently model-dependent manner.

  \item \textbf{Open issues and future outlook:} 
  The density and isospin dependence of SRCs above saturation density remain largely unconstrained, leaving open questions regarding their roles in NS cores and inner crusts, exotic degrees of freedom, in-medium pion and $\rho$-meson dynamics, and higher-order EOS coefficients; addressing these challenges requires coordinated advances in \emph{ab initio} nuclear many-body theory, heavy-ion reaction experiments, and multimessenger astrophysics.
\end{itemize}

\section*{Acknowledgment}

\indent 

We would like to thank Rui Wang for helpful discussions.
This work was supported in part by the National Natural Science Foundation
of China under contract No. 12547102, the U.S. Department of Energy, Office of Science, under Award Number DE-SC0013702, the CUSTIPEN (China-U.S. Theory Institute for Physics with
Exotic Nuclei) under the US Department of Energy Grant No. DE-SC0009971.

\bibliographystyle{unsrt}


{\footnotesize
\begin{spacing}{.85}
\balance
\bibliography{Ref}
\end{spacing}
}

\end{document}